\documentclass{article}

\usepackage{geometry}
\usepackage{amsmath}
\usepackage{algorithm}
\usepackage{cleveref}
\usepackage{url}

\usepackage{cite}  
\usepackage{amssymb}
\usepackage{booktabs}
\usepackage{enumerate}
\usepackage{graphicx}
\usepackage{rotating}
\usepackage[font={footnotesize,it}]{caption}
\usepackage{subcaption}
\usepackage{multicol}
\usepackage{multirow}
\usepackage{algorithmicx}
\usepackage[noend]{algpseudocode}
\usepackage{listings}
\usepackage{tikz}
\usepackage{standalone}
\usepackage{bm}
\usepackage{bbm}
\usepackage{float}
\usepackage{flafter}
\usepackage{makecell}
\usepackage{siunitx}
\newcommand{\numINT}[1]{\num[group-separator = {,},group-minimum-digits = 4]{#1}}
\newcommand{\percent}[1]{\SI{#1}{\percent}}
\usepackage{soul}
\usepackage{xspace}
\usepackage{grffile} 
\usepackage{autonum}

\Crefname{ALC@unique}{Line}{Lines} 

\newboolean{SHOWEM}
\setboolean{SHOWEM}{false}  
\ifthenelse{\boolean{SHOWEM}}
{\usepackage[textsize=footnotesize,textwidth=3.5cm]{todonotes}}
{\usepackage[textsize=footnotesize,textwidth=3.5cm,disable]{todonotes}}
\newcommand{\todoNK}[1]%
{\todo[color=green!20,inline]{{\bf Nils:} #1}}
\newcommand{\todoMM}[1]%
{\todo[color=yellow!20,inline]{{\bf Marcus:} #1}}
\newcommand{\todoMMfly}[1]%
{\todo[color=yellow!20]{{\bf Marcus:} #1}}
\newcommand{\todoSE}[1]%
{\todo[color=red!20,inline]{{\bf Sebastian:} #1}}
\newcommand{\todURi}[1]%
{\todo[color=blue!10,inline]{{\bf Uli:} #1}}
\newcommand{\todUR}[1]%
{\todo[color=blue!10,author=Uli,linecolor=darkgray]{#1}\xspace}

\crefname{equation}{}{}

\floatstyle{plaintop}
\restylefloat{table}

\usetikzlibrary{calc}
\usetikzlibrary{decorations.markings}
\usetikzlibrary{shapes,snakes}
\usetikzlibrary{shapes.geometric}

\newtheorem{remark}{Remark}[subsection]

\newcommand{\ie}{\mbox{i.\,e.}\xspace}
\newcommand{\eg}{\mbox{e.\,g.}\xspace}

\newcommand{\norm}[1]{\left\lVert#1\right\rVert}

\newcommand\restr[2]{{
  \left.\kern-\nulldelimiterspace 
  #1 
  \vphantom{\big|} 
  \right|_{#2} 
}}

\newcommand{\Pone}{$\mathbb{P}_1$\xspace}
\newcommand{\Ptwo}{$\mathbb{P}_2$\xspace}

\usepackage[acronym]{glossaries}
\makeglossaries

\newacronym{walberla}{\textsc{waLBerla}}{widely applicable Lattice-Boltzmann from Erlangen}
\newacronym{hhg}{HHG}{hierarchical hybrid grids}
\newacronym{hyteg}{\textsc{HyTeG}}{Hybrid Tetrahedral Grids}
\newacronym{mesapd}{\textsc{MESA-PD}}{Modular and Extensible Software Architecture for Particle Dynamics}
\newacronym{tme}{TME}{textbook multigrid efficiency}
\newacronym[plural=PDEs]{pde}{PDE}{partial differential equation}
\newacronym{pspg}{PSPG}{Pressure Stabilized Petrov-Galerkin}
\newacronym{fmg}{FMG}{full multigrid}
\newacronym{flops}{FLOPS}{floating point operations per second}
\newacronym{wu}{WU}{work unit}
\newacronym{mumps}{MUMPS}{MUltifrontal Massively Parallel sparse direct Solver}
\newacronym{petsc}{PETSc}{Portable, Extensible Toolkit for Scientific Computation}
\newacronym{ecm}{ECM}{Execution-Cache-Memory}
\newacronym{cfl}{CFL}{Courant-Friedrichs-Lewy}
\newacronym[shortplural=DoFs, longplural={degrees of freedom}]{dof}{DoF}{degree of freedom}
\newacronym{mmoc}{MMOC}{\emph{modified method of characteristics}}
\newacronym{ellam}{ELLAM}{Eulerian-Lagrangian localized adjoint method}
\newacronym{supg}{SUPG}{streamline upwind Petrov-Galerkin}
\newacronym{afc}{AFC}{algebraic flux correction}
\newacronym{fct}{FCT}{flux-corrected transport}
\newacronym{rk}{RK}{Runge-Kutta}
\newacronym{rhs}{RHS}{right-hand side}
\newacronym{cfd}{CFD}{computational fluid dynamics}
\newacronym[shortplural=ELMs, longplural={\emph{Eulerian-Lagrangian methods}}]{elm}{ELM}{\emph{Eulerian-Lagrangian method}}

\newcommand{\supermucng}{\mbox{SuperMUC-NG}\xspace}

\begin{document}

\title{A massively parallel Eulerian-Lagrangian method for advection-dominated
transport\\in viscous fluids}
\author{%
Nils Kohl\thanks{Computer Science 10, Friedrich-Alexander-Universit\"at Erlangen-N\"urnberg (\protect\url{nils.kohl@fau.de},
\mbox{\protect\url{ulrich.ruede@fau.de}}, \mbox{\protect\url{sebastian.eibl@fau.de}}).} \and
Marcus Mohr\thanks{Dept.~of Earth and Environmental Sciences, LMU Munich (\protect\url{marcus.mohr@lmu.de}).} \and
Sebastian Eibl\footnotemark[1] \and
Ulrich R{\"u}de\footnotemark[1] \thanks{Centre Europ\'een de Recherche et de Formation Avanc\'ee en Calcul Scientifique (CERFACS), France}}
\date{}
\maketitle

\begin{abstract}
    Motivated by challenges in Earth mantle convection,
    we present a massively parallel implementation of an Eulerian-Lagrangian method 
    for the advection-diffusion equation in the advection-dominated regime.
    The advection term is treated by a particle-based, characteristics method
    coupled to a block-structured finite-element framework.
    Its numerical and computational performance is evaluated in multiple,
    two- and three-dimensional benchmarks, including curved geometries,
    discontinuous solutions, pure advection, and it is applied to a coupled 
    non-linear system modeling buoyancy-driven convection in Stokes flow.
    We demonstrate the parallel performance in a strong and weak scaling experiment,
    with scalability to up to $147,456$ parallel processes, solving for more than
    $5.2 \times 10^{10}$ (52 billion) degrees of freedom per time-step.
\end{abstract}

\paragraph*{Key words}
Eulerian–Lagrangian methods, advection-diffusion, parallel algorithms

\paragraph*{AMS subject classifications}
65M25, 65Y05, 65M60

\glsresetall


\section{Introduction}\label{sec:introduction}
While to us as human beings the ground on which we walk may appear 'rock-solid'
the surface of our planet is actually in constant albeit very slow motion. 
Continental plates move at a rate of centimetres per year. The reason for this
movement are enormous forces acting deep below our feet. Convective processes
in the Earth's mantle help the planet rid itself of excess energy that is
either left from the time of its formation or generated by continued
radioactive decay. The mantle is a layer
of Earth starting from below the crust at roughly
\SI[group-separator = {,},group-minimum-digits = 4]{60}{km} and extending down
to the core-mantle-boundary at a depth of about 
\SI[group-separator = {,},group-minimum-digits = 4]{3000}{km}. On geologic
time-scales the rocks inside the mantle behave
like a highly viscous fluid.
A single overturn of the material in the mantle takes
about 100 mio.~years.

A detailed understanding of these processes is of fundamental interest to
geophysics, as they are the driving force behind phenomena such as
plate tectonics, mountain and ocean building, volcanism, and finally
earthquakes. As the mantle is not accessible for direct measurements studies
of its convection rely mostly on simulation and form an active research topic
in \gls*{cfd}. The requirements on spatial and temporal resolution render the
solution of the underlying system of \glspl*{pde} a grand challenge in
computational science \cite{Burstedde:2013:GJI,Bauer:2020:SPPEXA}.

The combination of extremely viscous material, characteristic length scale,
and creeping flow of the Earth's mantle result in a Reynolds number on the
order of $10^{-15}$, \cite{Ricard:2007:Treatise} and the Stokes equations
are suitable to model momentum and mass balance. Conservation of energy can
be described by an equation of advection--diffusion type for the temperature.
In a buoyancy-driven flow the dimensionless Rayleigh number Ra describes the
vigor of convection. For the Earth's mantle Ra is estimated to lie between
$10^7$ and $10^8$ \cite{Ricard:2007:Treatise}. In that range temperature 
transport is mainly driven by fluid flow (advection) and much less by diffusive
effects.

In this paper we are interested in the numerical treatment of this kind of
equation in the advection-dominated regime. While the temperature equation
of mantle convection forms our focus point, such kind of transport problems
appear, of course, also in many other applications in \gls*{cfd}
\cite{Chen:2006:SIAM,Morton:2019:CRC}. 
Although the quantity of interest varies,
the main characteristics of the underlying equation remain the same.
Typical transported variables include for example
chemical species concentration, material markers, or isotope ratios.

The solution of the advection-diffusion equation is known to be challenging in
the advection-dominated regime, for instance due to stability issues at high
gradients or even discontinuities in the solution
\cite{Quarteroni:2008:Springer,Elman:2014:OUP}. 
Well-known and established methods for the numerical treatment of
advection-diffusion equations include the \gls*{supg} method 
\cite{Brooks:1982:CMAME}, where for stability
reasons, artificial diffusion is introduced into the solution. 
A more recent approach in the
same direction is the entropy viscosity method, see
e.g.~\cite{Kronbichler:2012:GJI} and references therein.
\Gls*{afc} approaches the problem by modification of the equations at the
algebraic level \cite{Kuzmin:2012:Springer}. A comparison of \gls*{supg},
\gls*{afc} and other finite-element based methods for advection-dominated
transport is presented in \cite{John:2008:CMAME}.
High-order, discontinuous Galerkin discretizations \cite{Cockburn:1998:SINUM,Reinarz:2020:CPC}
are attractive as they are naturally well-suited to represent discontinuous
solutions. However, the selection of adequate slope-limiters and the large
number of unknowns that are introduced may be problematic.

A fundamentally different approach to the discretization of advection-diffusion
equations are so-called \emph{Lagrangian} or \emph{characteristic} methods.
Instead of employing a fixed, \emph{Eulerian} grid, the advected property is
captured by particles or volumes that move along the characteristics of the
velocity field. Usually, both, Eulerian and Lagrangian discretization
approaches are combined by means of a splitting-technique, where the advective
term is treated by a Lagrangian, and the diffusive term by an Eulerian
discretization. Solutions need to be interpolated between these two domains.
These approaches are also called \glspl*{elm}. Two prominent implementations of
this category are the \gls*{mmoc}
\cite{Douglas:1982:SINUM,Allievi:2000:IJNMF,Malevsky:1991:PFA,ElAmrani:2008:IJCM} 
(also referred to as characteristic Galerkin method or Lagrange-Galerkin
method) and the \gls*{ellam} \cite{Celia:1990:AWR,Russell:2002:AWR}. 

The \gls*{mmoc} is based on the backtracking of particles along the
characteristics, where the transported quantity for the next time step is
evaluated. This method permits large time steps, is free from parameterization
and conceptually easy to understand. 
The particle-based method requires frequent evaluation (or interpolation) of 
the solution function away from the grid nodes. In general, the \gls*{mmoc} 
is not perfectly energy-conserving. A scheme to enforce global energy 
conservation is developed in \cite{Douglas:1999:NUMA}.
Numerical analysis on accuracy and stability of the \gls*{mmoc} is found in 
\cite{Dawson:1989:SINUM,Bermudez:2006pt1:SINUM}.
Note that by following characteristics backwards in
time, \gls*{mmoc} is conceptually different from the particle/marker-in-cell
techniques often employed in geodynamical flow simulations for advecting
quantities like chemical composition or water content,
\cite{Gassmoeller:2019:GJI}. It also avoids some of their pitfalls such as
e.g.~the question of particle concentration per cell. The only investigation
of \gls*{mmoc}-based methods for geodynamical flows seems to be
\cite{Malevsky:1991:PFA}.

\gls*{ellam} may provide local energy conservation by propagation of volumes
instead of particles. This class of methods has similar advantages as the
\gls*{mmoc}, but the integration over elements that are not aligned with the
grid may be difficult, in particular in parallel implementations, and thus it
can be computationally expensive.

In this article, our focus is on an \gls*{elm} based on the \gls*{mmoc} suited
for massively parallel simulations on state-of-the-art supercomputers.
The parallel algorithms and data structures used in our implementation build 
upon the concept of \gls*{hhg} \cite{Bergen:2004:NLAA,Bauer:2020:SPPEXA}, 
addressing extreme-scalable, matrix-free geometric multigrid solvers on
block-structured grids. With
mantle convection models as a target application, a prototype application has
demonstrated scalability of Stokes solvers for systems with more than $10^{13}$
unknowns \cite{Gmeiner:2016:JoCS}. New matrix-free methods
\cite{Bauer:2017:ANM,Bauer:2018:SISC}, performance and scalability 
\cite{Gmeiner:2015:SISC,Gmeiner:2016:JoCS,Kohl:2020:arXiv},
and application to geophysical problems \cite{Bauer:2019:JoCS} have been
studied, mainly focusing on the solution of the Stokes system.
The \gls*{elm} proposed in this article is developed to exploit and extend
the excellent scalability of the \gls*{hhg}-based solvers for time-dependent
mantle-convection problems.

Parallel implementations of \glspl*{elm} have been designed for various
applications, including research on sea-ice \cite{Samake:2017:JCP},
Navier-Stokes \cite{Ouro:2019:CAF,Tavelli:2019:IJNMF}, and also natural
convection in \cite{Busto:2020:CAF}. In the latter a target application similar
to this work is considered on unstructured meshes, and an \gls*{elm}
is used for both the advection terms in the energy equation and also the
discretization of the Navier-Stokes system itself. 
However, only moderate scalability with up to \numINT{1000} parallel processes
was demonstrated. To quantitatively and accurately predict the convection
patterns of Earth's mantle, however, extreme-scale parallel simulations are
necessary, as for instance a global spatial resolution of $\sim$ 1.7km results
in linear systems with more than a trillion ($10^{12}$) \glspl*{dof}
\cite{Bauer:2019:JoCS}.
Such problems require methods that can efficiently exploit the resources of
today's peta- and future exascale supercomputers. With the proposed method,
we demonstrate the scalability of \gls*{elm}-based time-dependent simulations
for up to a hundred of thousand parallel processors.

\paragraph{Contribution}

In this paper we will

(a) present a particle-based, massively parallel method for the
advection-diffusion equation based on the \gls*{mmoc} that is applicable to
curved geometries and largely independent of the underlying grid data
structures and spatial discretization,

(b) embed the method into to a block-structured finite-element framework based
on \gls*{hhg},

(c) quantify the accuracy and energy conservation of our approach
through multiple, two- and three-dimen\-sio\-nal benchmarks with different spatial finite-element
discretizations, discontinuous solutions, pure advection, curved domains, large
time steps, \gls*{cfl} number $ > 1$, and coupled buoyancy-driven flow, and
	
(d) demonstrate the extreme-scalability of the approach on to up to
\numINT{147456} parallel processes and more than \num{5.2e10} particles,
and an application to a simplified mantle convection setup.

\paragraph{Reproducibility}

All presented algorithms and benchmarks are implemented in the open-source
software framework \gls*{hyteg}\footnote{\url{https://i10git.cs.fau.de/hyteg/hyteg}}
\cite{Kohl:2019:IJPEDS,Kohl:2020:arXiv,HyTeG:2021:SW},
assuring reproducibility of the results.

\subsection*{Governing equations}

We consider the numerical approximation of the advection-diffusion
equation on a bounded domain $\Omega \subset \mathbb{R}^d,\ d \in \{2,3\}$,
and time interval $[0, T], T \in \mathbb{R}^+$
\begin{align}\label{eq:advection-diffusion-pde}
	\frac{\partial}{\partial t} c  + \mathbf{u} \cdot \nabla c - \kappa \Delta c = q,
	&\quad (\mathbf{x}, t) \in \Omega \times [0, T]
\end{align}
where $c = c(\mathbf{x}, t)$ represents the advected, scalar 
quantity (temperature in case of our target application), 
$\mathbf{u} = \mathbf{u}(\mathbf{x}, t)$ a given
divergence-free velocity field, \ie satisfying
\begin{align}
	\nabla \cdot \mathbf{u} = 0, 
	\quad (\mathbf{x}, t) \in \Omega \times [0, T]\enspace,
\end{align}
$q = q(\mathbf{x}, t)$ the given rate of internal heat production, 
and $\kappa \geq 0$ a diffusivity parameter. 
Initial, Dirichlet, and (homogeneous) Neumann boundary conditions for the temperature $c$ are given by
\begin{equation}
  c(\mathbf{x}, 0) = c_0(\mathbf{x}),\ \mathbf{x} \in \Omega, \quad
  c(\mathbf{x}, t) = c_{\Gamma}(\mathbf{x}, t),\ \mathbf{x} \in \partial \Omega_D, \quad
  \frac{\partial c}{\partial \mathbf{n}}(\mathbf{x}, t) = 0,\ \mathbf{x} \in \partial \Omega_N
\end{equation}
for $t \in [0, T]$, boundary $\partial \Omega = \partial \Omega_D \cup \partial \Omega_N$, and
outward normal $\mathbf{n}$. We require for the sake of simplicity that the velocity field has no inflow into the domain.

In typical applications, the advective term 
$\mathbf{u} \cdot \nabla c$ strongly dominates over the diffusive term $\kappa \Delta c$.
Depending on the formulation and non-dimensionalization of the model, this translates 
to either $\kappa \ll 1$, or large velocity magnitudes.

The advection-diffusion equation can be coupled to the Stokes equation for viscous flows
using the Boussinesq-approximation for natural convection, as will be described in \cref{sec:coupled-flow}.


\section{Eulerian-Lagrangian method}

In this section we describe the parallel algorithms and data structures of the \gls*{mmoc}-based method 
for the advection-diffusion equation \cref{eq:advection-diffusion-pde}.

\subsection{Hierarchical hybrid grids}\label{sec:domain}

We base the construction of the computational mesh on the concept of 
\gls*{hhg} \cite{Bergen:2004:NLAA,Bauer:2020:SPPEXA}. Therefore, we define
a coarse unstructured mesh $\mathcal{T}_0$ of  tetrahedral (or triangular) elements that partitions
the domain $\Omega$. In a second step, each coarse grid element is uniformly
refined according to \cite{Bey:1995:Tetrahedral}. This results in hierarchy of
block-structured meshes $\mathcal{T} = \{\mathcal{T}_\ell,\, \ell = 0, ..., L\}$
and 
offers crucial performance advantages for matrix-free multigrid methods as demonstrated
especially for the Stokes system 
\cite{Kohl:2020:arXiv,Bauer:2020:SPPEXA,Bauer:2017:ANM,Bauer:2018:SISC}.

If the problem domain $\Omega$ is polyhedral, we can define a set of coarse grid elements,
whose union equals $\Omega$. However, in this article we also consider a more general case,
which is that $\Omega$ coincides with a polyhedral domain after a \emph{blending function}
$\Phi$ is applied to the latter.
In particular, we are interested in domains with curved boundaries, such as the thick
spherical shell, that is used to represent Earth's mantle in geophysical models
\cite{Bauer:2019:JoCS,Rudi:2015:SC}. We require $\Phi$ to be a homeomorphism and its
inverse to be known explicitly.

\begin{figure}[!ht]
    \footnotesize
    \centering
    \begin{subfigure}[t]{0.48\textwidth}
		\centering
		\includegraphics[width=0.8\textwidth]{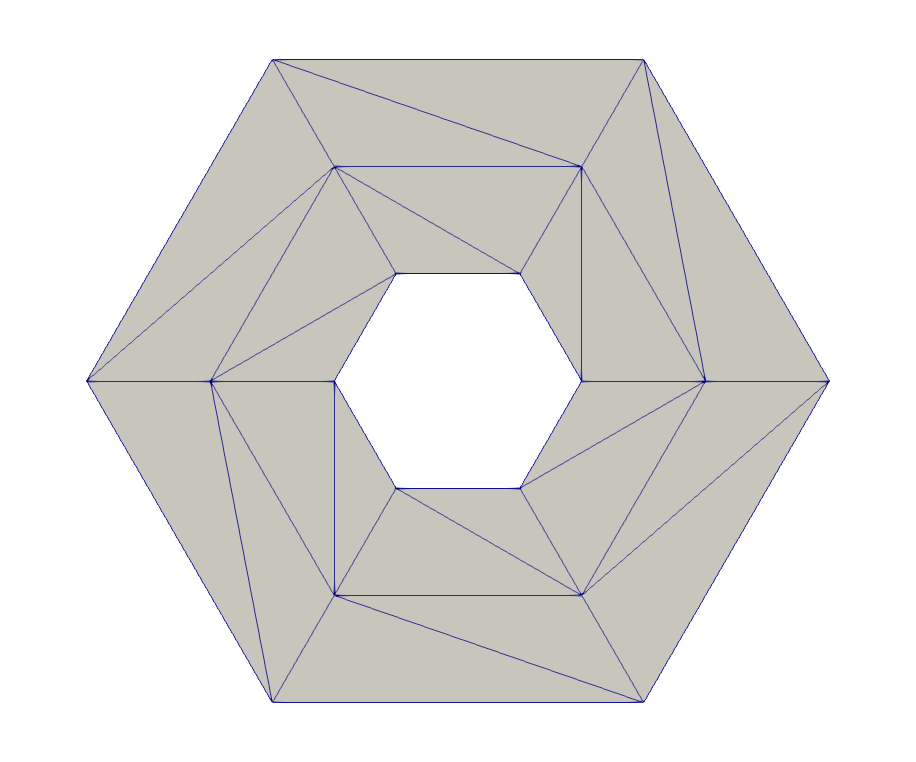}
    	\caption{$\mathcal{T}_0$}
		\label{fig:annulus-domain-comp}
	\end{subfigure}
    \hfill
	\begin{subfigure}[t]{0.48\textwidth}
		\centering
		\includegraphics[width=0.8\textwidth]{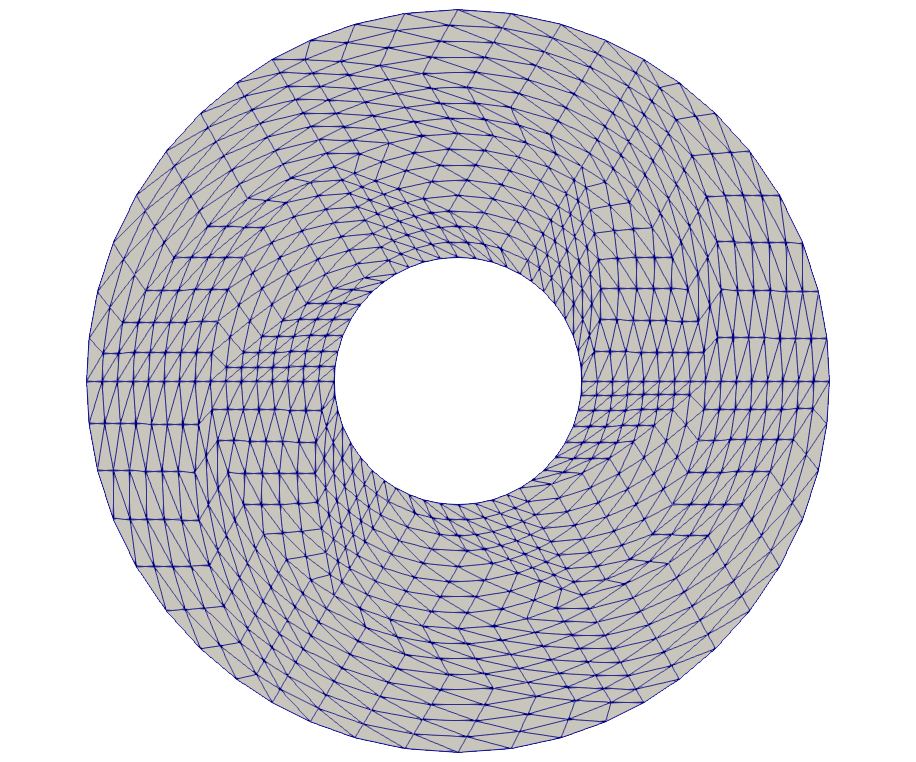}
    	\caption{$\Phi(\mathcal{T}_3)$}
		\label{fig:annulus-domain-phy}
	\end{subfigure}
	\footnotesize
    \caption{Partitioning of an annular domain:
	(\protect\subref{fig:annulus-domain-comp}) unstructured, initial computational mesh before refinement,
	(\protect\subref{fig:annulus-domain-phy}) refined mesh mapped to physical domain.}
    \label{fig:annulus-domain}
\end{figure}

To construct the grid hierarchy for this second case,  
we start from an approximation of the 
\emph{physical domain} $\Omega_\text{phy} := \Omega$ by a polyhedral, \emph{computational domain}
$\Omega_\text{comp}$ (\ie $\Phi(\Omega_\text{comp}) = \Omega_\text{phy}$). 
This polyhedral domain is then refined as outlined above, yielding
a mesh hierarchy $\mathcal{T} = \{\mathcal{T}_\ell,\, \ell = 0, ..., L\}$.
Finally, by applying our blending function to each mesh $\mathcal{T}_\ell$ we
obtain a hierarchy $\widetilde{\mathcal{T}} := \{\Phi(\mathcal{T}_\ell),\, \ell = 0,
\ldots, L\}$ for $\Omega_\text{phy}$. Obviously, application of this
algorithm to a polyhedral physical domain $\Omega_\text{phy}$ corresponds to the special case
$\Phi = \mathrm{Id}$ as $\Omega_\text{phy} = \Omega_\text{comp}$.
\Cref{fig:annulus-domain} shows an example, where the computational domain is
projected onto an annulus. The left figure shows an initial, unrefined, unstructured computational
mesh $\mathcal{T}_0$, the right figure the corresponding physical mesh $\Phi(\mathcal{T}_3)$ after 
three refinement iterations.

Efficient and scalable, matrix-free solvers for scalar elliptic PDE problems
and Stokes flow on curved domains
in conjunction with \gls*{hhg} have been presented in \cite{Bauer:2018:SISC,Bauer:2017:ANM}. 

\subsection{Discretization of the advection-diffusion equation}\label{sec:discretization}

The essence of the \gls*{mmoc} is the elimination of the advective term $\mathbf{u} \cdot \nabla c$
from \cref{eq:advection-diffusion-pde}. 
For this, we define the so-called 
\emph{characteristics} $\mathbf{X} : \Omega \times [0, T]^2 \rightarrow \mathbb{R}^3$
of the velocity field $\mathbf{u}$ as the solutions of
\begin{equation}\label{eq:characteristic-curves}
	\begin{aligned}
		\frac{d}{dt} \mathbf{X}(\mathbf{x}, s, t) &= \mathbf{u}(\mathbf{X}(\mathbf{x}, s, t), t), 
		\quad t \in (0, T) \\
		\mathbf{X}(\mathbf{x}, s, s) &= \mathbf{x}
	\end{aligned}
\end{equation}
for fixed $(\mathbf{x}, s) \in \Omega \times [0, T]$.
Specifying two points in time $t_0, t_1 \in [0, T],\, t_0 < t_1$, 
$\mathbf{X}(\mathbf{x}, t_1, t_0)$ can be interpreted as the \emph{departure point} 
at time $t_0$ of a particle, that reaches point $\mathbf{x}$ at time $t_1$.
Such a departure point is, thus, given by
\begin{align}\label{eq:departure-point}
	\mathbf{X}(\mathbf{x}, t_1, t_0) = \mathbf{x} - \int_{t_0}^{t_1} \mathbf{u}(\mathbf{X}(\mathbf{x}, t_1, t), t) \, dt.
\end{align}
We now define, for a fixed time $s \in [0, T]$
\begin{align}
	\hat{c}(\mathbf{x}, t) := c(\mathbf{X}(\mathbf{x}, s, t), t)
\end{align}
and calculate, using the chain rule and \cref{eq:characteristic-curves}
\begin{align}\label{eq:material-derivative}
	\frac{\partial}{\partial t} \hat{c} (\mathbf{x}, t) = 
	\left( \frac{\partial}{\partial t} c  + 
	\mathbf{u} \cdot \nabla c \right) (\mathbf{X}(\mathbf{x}, s, t), t).
\end{align}
At time $t = s$ 
we can replace the advective term
in \cref{eq:advection-diffusion-pde}, since
\begin{align}\label{eq:material-derivative-t-eq-s}
  \frac{\partial}{\partial t} \hat{c} (\mathbf{x}, s) = 
  \left( \frac{\partial}{\partial t} c  + 
  \mathbf{u} \cdot \nabla c \right) (\mathbf{x}, s),
\end{align}
and reformulate the PDE as
\begin{align}\label{eq:advection-diffusion-pde-reformulated}
	\frac{\partial}{\partial t} \hat{c} - \kappa \Delta c = q.
\end{align}
Next we semi-discretize \cref{eq:advection-diffusion-pde-reformulated} in time.
To this end, we divide $[0, T]$ into $N$ intervals $[t_n, t_{n+1}]$, $n \in \{0, \dots, N-1\}$ with step size
$\tau_n = t_{n+1} - t_n$.
We then set $\mathbf{x} = \mathbf{X}(\mathbf{x}, t_{n+1}, t_{n+1})$ (or $s = t_{n+1}$
in \cref{eq:material-derivative}) and approximate the time derivative via a difference quotient
\begin{align}\label{eq:mmoc-approximation}
	\frac{\partial}{\partial t}\hat{c}(\mathbf{x}, t_{n+1})  \approx
	\frac{1}{\tau_n} \Big[ \hat{c}(\mathbf{x}, t_{n+1}) - \hat{c}(\mathbf{x}, t_{n}) \Big] 
	= \frac{1}{\tau_n} \Big[ c(\mathbf{x}, t_{n+1}) - c(\mathbf{X}(\mathbf{x}, t_{n+1}, t_{n}), t_{n}) \Big].
\end{align}

We perform the spatial discretization of the temperature and velocity fields using
the standard Galerkin finite element method subject to the \gls*{hhg} grid hierarchy described
in \cref{sec:domain}. 
We therefore introduce the spaces of piecewise polynomial functions
\begin{align}
\mathcal{S}_\ell^{m} := \{ v \in \mathcal{C}^0(\Omega) : \restr{v}{T} \in \mathcal{P}_m(T),\ \forall \ T \in \mathcal{T}_\ell \}, \quad l \in \{0, ..., L\}, \ m \geq 1.
\end{align}
Here, $\mathcal{P}_m(T)$ denotes the space of polynomials of degree $m$ on the element $T$.
Let $V_h := \mathcal{S}_{L}^{m} \cap \mathcal{H}^1_0(\Omega)$ be a finite 
dimensional subspace of $\mathcal{H}^1_0(\Omega)$ with piecewise polynomial basis functions that vanish on the boundary.
In particular, we employ the standard sets of Lagrange basis functions $P_m$ for polynomial
degree $m$ \cite{Elman:2014:OUP}.
Furthermore, given a function $c_{\Gamma} := c_{\Gamma}(\mathbf{x}, t)$ 
that defines suitable Dirichlet boundary conditions, let 
$V^D_h := \mathcal{S}_{L}^{m} \cap \mathcal{H}^1_D(\Omega)$ with
$\mathcal{H}^1_D := \{ v_h \in \mathcal{H}^1(\Omega) : v_h = c_{\Gamma} \text{ on } \partial \Omega_D \}$.

We apply the $\Theta$-method to the time-discretization of the diffusive term \cite{Quarteroni:2008:Springer}.
The finite dimensional
version of the weak formulation of \cref{eq:advection-diffusion-pde-reformulated} then reads: given 
$\hat{c}_h^{n} = \hat{c}_h^{n}(\mathbf{x}) \in V^D_h$, find $c_h^{n+1} = c_h^{n+1}(\mathbf{x}) \in V^D_h$ so that
\begin{equation}\label{eq:finite-dimensional-galerkin-approximation}
	\begin{aligned}
	\frac{1}{\tau_n} (c_h^{n+1} - \hat{c}_h^{n}, v_h) &+ \Theta \kappa (\nabla c_h^{n+1}, \nabla v_h) + (1 - \Theta) \kappa (\nabla \hat{c}_h^{n}, \nabla v_h) \\
	&= (\Theta q(t_{n+1}) + (1 - \Theta) q(t_n), v_h), \quad \text{for all } v_h \in V_h 
	\end{aligned}
\end{equation}
and $c_h^{0} = c_{0,h}$. $(\cdot, \cdot)$ denotes the inner product in $L^2(\Omega)$ and $\Theta \in [0, 1]$. 
This corresponds to an implicit Euler or Crank-Nicolson scheme for the diffusive term,
for $\Theta = 1$ or $\Theta = 0.5$, respectively.
For the formulation of the bilinear and linear forms in the case of a blended domain, 
\ie $\Phi \neq \mathrm{Id}$, we refer to \cite{Bauer:2018:SISC,Gordon:1973:NUMA}.

Associating $c_h^{n+1}$, $\hat{c}_h^n$, $(q(t_n), v_h)$, and $(q(t_{n+1}), v_h)$ with coefficient vectors $\underline{\mathbf{c}}^{n+1}$,
$\underline{\mathbf{\hat{c}}}^{n}$, $\underline{\mathbf{q}}^{n}$, and $\underline{\mathbf{q}}^{n+1}$
we formulate \cref{eq:finite-dimensional-galerkin-approximation} as the linear system
\begin{equation}\label{eq:linear-system}
	\begin{aligned}
	(M + \tau_n \Theta \kappa A) \underline{\mathbf{c}}^{n+1} ={} (M - \tau_n(1 - \Theta) \kappa A)\underline{\mathbf{\hat{c}}}^{n} 
	+ \tau_n (\Theta \underline{\mathbf{q}}^{n+1} + (1-\Theta) \underline{\mathbf{q}}^{n})
	\end{aligned}
\end{equation}
that has to be solved in each time step. 
$M$ represents the finite element mass matrix, and
$A$ the stiffness matrix.
The matrix $E := (M + \tau_n \Theta \kappa A)$ is symmetric and positive definite. This allows for efficient
inversion. Especially for small time steps, $E$ tends to be more diagonally dominant than
the stiffness matrix $A$ and is therefore well suited for treatment with conjugate gradient and
multigrid solvers \cite{Trottenberg:2001:GreyBook}. 

It remains to determine an approximation for $\underline{\mathbf{\hat{c}}}^{n}$, which requires
the evaluation of $\hat{c}^n_h(\mathbf{x}) = c^n_h(\mathbf{X}(\mathbf{x}, t_{n+1}, t_{n}))$.
The advected temperature is obtained by calculation of the departure point 
$\mathbf{X}(\mathbf{x}, t_{n+1}, t_{n})$ via the integral in \cref{eq:departure-point}.
Due to the initial condition and the continuous Galerkin discretization, 
$c^n_h(\mathbf{x})$ can be evaluated for all $\mathbf{x} \in \Omega$.

In general, the integral in \cref{eq:departure-point} cannot be evaluated analytically
but has to be approximated numerically. Here, we apply standard, explicit \gls*{rk} schemes
that repeatedly evaluate the velocity field $\mathbf{u}$. For the general case of time-dependent and
time-discrete velocity fields, evaluation at time $t^* \in (t_n, t_{n+1})$ requires
interpolation. In this case, we employ linear interpolation in time.
Spatially, we represent the velocity field $\mathbf{u}$ also in one of the continuous finite element
spaces $\mathcal{S}_\ell^{m}$
resulting in a well-defined approximation $\mathbf{u}_h$. Details on the numerical integration and
evaluation are presented in \cref{sec:implementation}.

\Cref{alg:ad} summarizes the time-stepping scheme for the advection-diffusion equation.
To determine a suitable time-step size, we employ a \gls*{cfl} condition via a constant 
$\text{CFL}_\text{max}$, the length of the shortest edge of the mesh $h_\text{min}$,
and the maximum velocity magnitude at time-step $n$, \ie $\max_{\mathbf{x}\in\Omega}|\mathbf{u}_h(\mathbf{x}, t_n)|$.
\begin{algorithm}
    \footnotesize
    \algloopdefx{Repeat}[1]{\textbf{repeat} #1 \textbf{times}}
    \begin{algorithmic}[1] 
		\Procedure{AD}{$c_h^n, \mathbf{u}_h$}
			\State $\tau_n = \text{CFL}_\text{max} \cdot h_\text{min} / \max_{\mathbf{x}\in\Omega}|\mathbf{u}_h(\mathbf{x}, t_n)|$  
				\Comment{determine time-step size}
			\State $\hat{\mathbf{x}} = \mathbf{X}(\mathbf{x}, t_{n+1}, t_{n})$ 
				\Comment{calculate departure points (see \cref{sec:implementation})}\label{alg:ad:departure-points}
			\State $\hat{c}_h^{n}(\mathbf{x}) = c_h^{n}(\hat{\mathbf{x}})$
				\Comment{advection}\label{alg:ad:advection}
			\State solve \cref{eq:linear-system} to advance from $\hat{c}_h^{n}$ to $c_h^{n+1}$
				\Comment{diffusion}
            \State \textbf{return} $c_h^{n+1}$
        \EndProcedure
    \end{algorithmic}
	\caption{\footnotesize Time-stepping scheme, advection-diffusion.}
    \label{alg:ad}
\end{algorithm}

\subsection{Parallel implementation}\label{sec:implementation}

In this section, we describe the parallel implementation of the \gls*{mmoc} on \gls*{hhg}.
In particular, we discuss the execution of the Lagrangian step, 
\ie the calculation of $\hat{c}^n_h$ ,
and the implementation in the \gls*{hyteg} finite element framework. 
This corresponds to lines~\ref{alg:ad:departure-points}, and~\ref{alg:ad:advection} in \cref{alg:ad}.

\subsubsection{Particle tracing}\label{sec:particle-tracing}

We employ \emph{tracer particles}
that are created at the \glspl*{dof} of $c_h$ at time $t_{n+1}$ and are transported backwards along the 
velocity trajectories, until they reach the departure points at time $t_n$.
Usually, for standard Lagrange finite element discretizations, the \glspl*{dof} are set to coincide
with the grid vertices for a \Pone discretization, and with the vertices and edge-midpoints
for a quadratic \Ptwo discretization. However, the method is not restricted to
such a choice, and discretizations with a different \gls*{dof}-layout such as finite-volumes may also be realized.
The values of $\hat{c}^n_h$ at the \glspl*{dof}
are then determined by evaluation of $c^n_h$ at the departure points.

Given the continuous Galerkin approximation $c^n_h$ of $c$ on the \gls*{hhg} structure, 
we split the approximation of $\hat{c}^n_h$ into three steps: 
(i) particle creation, (ii) particle integration and (iii)
temperature evaluation.
It follows a discussion of the grid and particle data structures, and steps
(i) -- (iii).

\paragraph{Grid data structure}
For each element of the unstructured coarse grid,
a \emph{macro-primitive} (macro-faces in 2D, macro-cells in 3D) 
data structure is created. The macro-primitives are then uniformly refined. 
The \gls*{hhg} concept introduces \emph{interface primitives}
for each interface between two coarse grid elements. 
The interface primitives are also refined uniformly. 
As an example, in 2D, two neighboring \emph{macro-face} primitives are interfaced by a 
\emph{macro-edge} primitive,
and two adjacent macro-edges are interfaced by a \emph{macro-vertex}.
This allows for a unique assignment of each individual \gls*{dof} to a single 
primitive data structure.
Each primitive is assigned a globally unique ID, and in a parallel setting,
assigned to one of the parallel processes. For distributed memory architectures,
communication is implemented via MPI. The coarse grid and all mesh-related metadata
are distributed without global data structures, allowing for parallel runs
on hundreds of thousands of parallel processes \cite{Gmeiner:2016:JoCS,Kohl:2020:arXiv}.
More details on the \gls*{hhg} data structures can be found in 
\cite{Bergen:2004:NLAA,Kohl:2019:IJPEDS,Kohl:2020:arXiv}.

\paragraph{Particle data structure and synchronization}

The tracer particles are realized by the \gls*{mesapd} \cite{Eibl:2018:PARCO,Eibl:2019:arXiv},
which implements particle data structures for massively parallel particle simulations.
It allows to equip each particle with arbitrary properties, that are transported together with the particle
through a distributed domain. The individual subdomains correspond to the volume primitives
defined by the unstructured coarse grid. Particles that leave the subdomain of
a process are communicated via MPI. Similar to the \gls*{hhg} structure, the parallel particle 
data structures are distributed to allow for massively parallel simulations
by design. 

After the position of a particle is updated, a synchronization step follows, that assigns
each particle uniquely to a single neighboring volume primitive. The target primitive is determined
only by the previous owner process of the particle, and therefore prevents race conditions.
Detailed information on the parallel data structures and communication are found in \cite{Eibl:2019:arXiv}.

\begin{figure}
    \footnotesize
    \centering
    \begin{subfigure}[t]{0.40\textwidth}
		\centering
		\includegraphics[width=\textwidth]{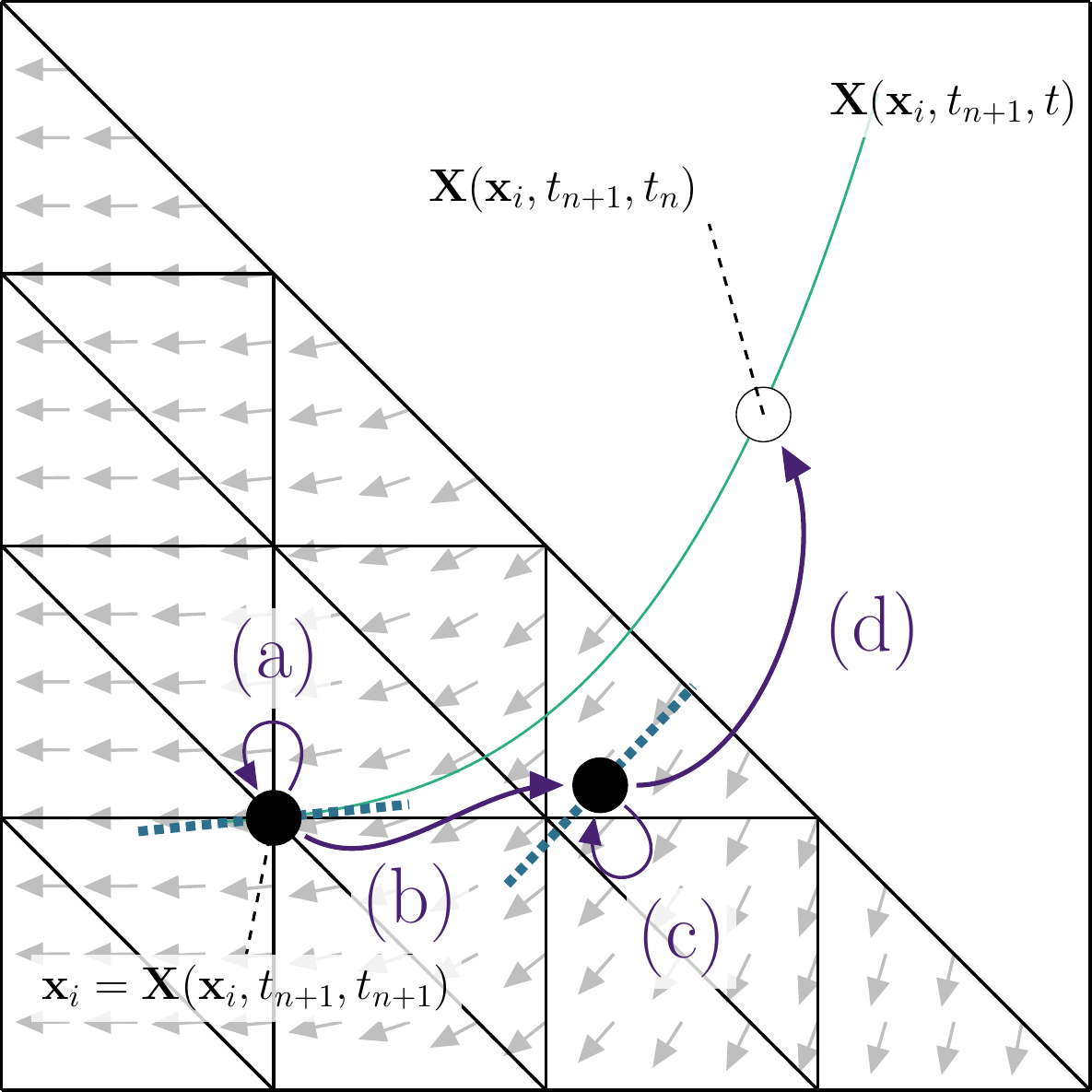}
	\end{subfigure}
	\hspace{12pt}
	\begin{subfigure}[t]{0.40\textwidth}
		\centering
		\includegraphics[width=\textwidth]{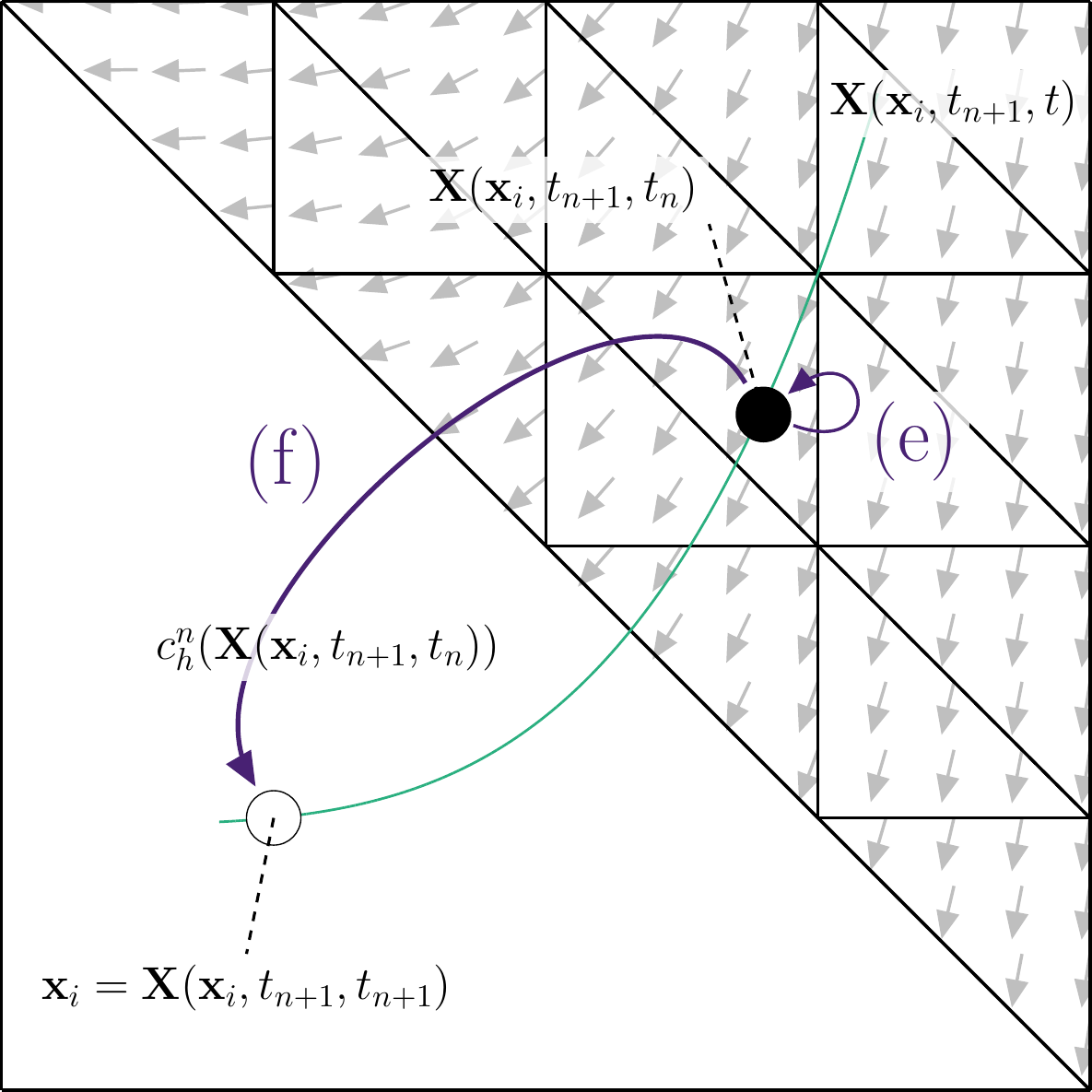}
	\end{subfigure}
    \caption{Illustration of the parallel particle integration and temperature
		evaluation (steps (ii) and (iii)) on two neighboring volume primitives. 
		In this example, a 2-stage RK method is employed: 
		(a) initial particle position, evaluation of 
		$\tilde{\mathbf{u}}^1_h(\mathbf{y}_i^1)$ (tangent to velocity field at that point, illustrated by dotted blue line), 
		(b) setting particle position to $\mathbf{y}^2_i$,
		(c) evaluation of 
		$\tilde{\mathbf{u}}^2_h(\mathbf{y}_i^2)$,
		(d) calculation of $\mathbf{X}(\mathbf{x}_i, t_{n+1}, t_n)$ using 
		$\tilde{\mathbf{u}}^1_h(\mathbf{y}_i^1)$ and 
		$\tilde{\mathbf{u}}^2_h(\mathbf{y}_i^2)$ according to the RK method,
		and particle communication to the neighboring volume primitive,
		(e) evaluation of $c^n_h(\mathbf{X}(\mathbf{x}_i, t_{n+1}, t_n))$, 
		(f) communication of $c^n_h(\mathbf{X}(\mathbf{x}_i, t_{n+1}, t_n))$
		back to initial \gls*{dof}.}
	\label{fig:particle-integration}
\end{figure}

\paragraph{Step (i): particle creation} For each \gls*{dof} of the Eulerian grid,
a particle is created.
The particles are initialized with the corresponding macro-primitive ID, \gls*{dof}-index,
and the process ID,
so that their corresponding \gls*{dof} can be backtracked in a distributed setting.
Particles are also initialized on
interface primitives, since they are responsible for \glspl*{dof} at the interfaces of
the volume primitives. The initial position of a particle corresponds to 
$\mathbf{x}_i = \mathbf{X}(\mathbf{x}_i, t_{n+1}, t_{n+1})$, where $\mathbf{x}_i$
is the location of a \gls*{dof} with index $i$ on $\Omega_\text{phy}$.
A following synchronization step assigns all particles that were created
on an interface primitive to a single volume primitive. 
It is of no particular importance which volume primitive is chosen.

\paragraph{Step (ii): particle integration}
This step performs the backward transport along the velocity field
using an explicit \gls*{rk} integrator with $S$ stages. This corresponds
to the computation of $\mathbf{X}(\mathbf{x}_i, t_{n+1}, t_n)$ according
to \cref{eq:departure-point} using numerical integration.
The RK integration requires the evaluation 
of the velocity field at a time $\tilde{t}^{s} \in [t_n, t_{n+1}]$ and position 
$\mathbf{y}_i^s$, with $\mathbf{y}_i^1 = \mathbf{x}_i$ in each stage $s \in [1, ..., S]$.

Before each RK stage, the position of a particle is set to the position 
$\mathbf{y}_i^s$ where the velocity field needs to be evaluated
(see \cref{fig:particle-integration} step (b)).
Immediately after that, a synchronization step follows so that all 
particles are available on the process that owns the volume-primitive that
contains $\mathbf{y}_i^s$.

We assume that the velocity field is known for the discrete time steps 
$t_n$ and $t_{n+1}$. Both fields $\mathbf{u}^n_h$ and $\mathbf{u}^{n+1}_h$ 
are evaluated and we perform linear interpolation. 
This means we approximate
\begin{align}\label{eq:linear-interpolation}
\mathbf{u}(\mathbf{y}_i^s, \tilde{t}^s) \approx \tilde{\mathbf{u}}^s_h(\mathbf{y}_i^s) :=
\left( \frac{t_{n+1} - \tilde{t}^s}{t_{n+1} - t_n} \right) \mathbf{u}^{n}_h(\mathbf{y}_i^s) + 
\left( \frac{\tilde{t}^s - t_n}{t_{n+1} - t_n} \right) \mathbf{u}^{n+1}_h(\mathbf{y}_i^s)
\end{align}
(see \cref{fig:particle-integration} steps (a) and (c)).
For scenarios where the velocity depends on the temperature field, we refer to \cref{sec:coupled-flow}
where we discuss buoyancy-driven flows.

The intermediate result $\tilde{\mathbf{u}}^s_h(\mathbf{y}_i^s)$ is stored in the
particle data structure before the next stage is executed. After the
last stage, all intermediate results and weights of the RK method are 
combined to calculate the actual final position 
$\mathbf{X}(\mathbf{x}_i, t_{n+1}, t_n)$
(see \cref{fig:particle-integration} step (d)).

\paragraph{Step (iii): temperature evaluation}
In this last step, the temperature field $c^n_h$ is evaluated at
$\mathbf{X}(\mathbf{x}_i, t_{n+1}, t_n)$ 
(see \cref{fig:particle-integration} step (e)).
This gives $\hat{c}^n_h(\mathbf{x}_i)$
at the initial position $\mathbf{x}_i$ of the particle. Since
the initial position was a \gls*{dof}, we set the corresponding coefficient
$\underline{\mathbf{\hat{c}}}^n_i = c^n_h(\mathbf{X}(\mathbf{x}_i, t_{n+1}, t_n))$.
If $\mathbf{x}_i$ is located on a different volume primitive than
$\mathbf{X}(\mathbf{x}_i, t_{n+1}, t_n)$, $\underline{\mathbf{\hat{c}}}^n_i$ 
is communicated (see \cref{fig:particle-integration} step (f)).

\subsubsection{Field evaluation}

The evaluation of 
$\mathbf{u}_h^n(\mathbf{z})$ or $c_h^n(\mathbf{z}),\ \mathbf{z} \in \Omega$ involves localization
of the underlying geometric element, and computing a sum of the shape functions evaluated at
$\mathbf{z}$ weighted by the corresponding DoF values.
In general, as described in \cref{sec:domain}, $\Omega = \Omega_\text{phy}$ may be non-polyhedral, 
\ie $\Phi \neq \mathrm{Id}$. 
We therefore map $\mathbf{z}$ to the computational domain $\Omega_\text{comp}$ and set 
$\mathbf{z}_\text{comp} := \Phi^{-1}(\mathbf{z})$. 
Since we require $\Phi$ to be a homeomorphism, we know that 
$\mathbf{z}_\text{comp} \in T \subset \Omega_\text{comp}
\Leftrightarrow \mathbf{z} \in \Phi(T) \subset \Omega_\text{phy}$.
We split the search-locate algorithm on the computational domain 
into two steps. In a first step, the enclosing 
volume-primitive that contains $\mathbf{z}_\text{comp}$ is determined by searching in the 
direct neighborhood of the volume-primitive that previously contained the corresponding particle. 
Then, we search for the containing element $T \subset \Omega_\text{comp}$ of $\mathbf{z}_\text{comp}$ 
in the uniformly refined volume primitive.
Since we employ block-structured \gls*{hhg}, the element $T$ is found in $\mathcal{O}(1)$ cost. 
Finally, the value of the finite element function is
computed as is standard, by application of a pull-back mapping of $T$ to the reference
element. 

\subsubsection{Look-back Distance}\label{sec:re-init}
The field evaluation in step (iii) implicitly corresponds to an interpolation of the
advected temperature field $\hat{c}^n_h$ into the space $\mathcal{S}_\ell^{m}$. While the
discretized original field 
at time $t_n$ satisfies $c^n_h\in\mathcal{S}_\ell^{m}$ this will typically
not be the case for $\hat{c}^n_h$. Consequently this step introduces an interpolation
error. If the field $\hat{c}^n_h$ used to update $c^{n+1}_h$ is computed from $c^n_h$, the latter
already involves $n$ previous interpolations, whose errors might accumulate.

However, in the purely advective case ($\kappa=0$ and $q = 0$)
this issue can be diminished or even completely removed. To do so, one can simple follow the
particle trajectory back in time over more than only one temperature time step $\tau$,
i.e.~instead of integrating from $t_{n+1}$ back to $t_n$ we select an earlier time
$t_{n+1-b}$. We will refer to the integer $b$ as \emph{look-back distance} as $t_{n+1-b}$
will be the time when temperature is evaluated. Of course, this approach requires that
the temperature field is still known at $t_{n+1-b}$, as must be the intermediate velocity
fields required by the ODE solver.

By selecting $b=n+1$ one can derive $c^{n+1}_h$ from the initial temperature $c_{0,h}$
itself. However, the look-back distance then grows with the simulation, a fact that we will
mark by using the notation $b=\infty$. This extreme approach preserves the accurate
representation of the initial temperature in the Lagrangian domain, and leads to very accurate
solutions, as we will see in the following benchmarks. There we will employ different look-back
distances $b$ to demonstrate that the interpolation between the Lagrangian and Eulerian
representation is the primary source of approximation error. A similar discussion of the
accumulation of the interpolation error is found in \cite{Malevsky:1991:PFA}.

\section{Numerical verification}\label{sec:benchmarks}

In the following subsections we assess the accuracy of our implementation through numerical benchmarks.

\subsection{Test setup}

In all benchmarks, we employ either linear (\Pone) or quadratic (\Ptwo) 
Lagrangian finite element discretizations
for the temperature and velocity, block-structured triangular and tetrahedral meshes for
two- and three-dimensional domains respectively. For the particle integration
we use the standard fourth-order \gls*{rk} integrator (often referred to as RK4).
We note that the implementation supports any explicit \gls*{rk} integrator.

To asses the quality of our scheme, we employ the following norms and metrics:
let $\tilde{c}_h$ the computed solution, $c_h$ the point-wise interpolated exact
solution and $e_h = c_h - \tilde{c}_h$. The corresponding coefficient vectors are
denoted as $\underline{\mathbf{c}}$, $\underline{\mathbf{\tilde{c}}}$, and $\underline{\mathbf{e}}$. 
A discrete version of the $\mathcal{H}^0$-norm of
the error is then defined as 
\begin{align}
    \norm{e_h}_{\mathcal{H}^0} := \left( \underline{\mathbf{e}}^\top M \underline{\mathbf{e}} \right)^\frac{1}{2}
\end{align}
where $M$ is the finite element mass matrix. 
We define $\text{var}(t_n)$ as in \cite{John:2008:CMAME}, and $E_\text{peak}(t_n)$ as
\begin{align}
    \text{var}(t_n) := \max_{j} \underline{\mathbf{\tilde{c}}}^n_j - \min_{j} \underline{\mathbf{\tilde{c}}}^n_j, \quad
    E_\text{peak}(t_n) := \frac{\max_{j} \underline{\mathbf{\tilde{c}}}^n_j}{\max_{j} \underline{\mathbf{c}}^n_j} - 1
\end{align}
to indicate the amount of spurious oscillations, and to detect if peaks of the solution are preserved.
To quantify the energy conservation of our implementation, we indicate a relative energy difference $\Delta m(t_n)$ compared to the initial solution by
\begin{align}
    \Delta m(t_n) := \frac{m(t_n)}{m(t_0)} - 1, \quad m(t_n) := \underline{\mathbf{1}}^\top M \underline{\mathbf{\tilde{c}}}^n, \quad
    \underline{\mathbf{1}} := (1, \dots, 1)^\top.
\end{align}

\subsection{Circular advection}

First we consider a two dimensional body rotation problem as employed in
\cite{Zalesak:1979:JCP,LeVeque:1996:SINUM,John:2008:CMAME}.
In particular, the setup is the same as in \cite{John:2008:CMAME} to compare the numerical results.

\begin{figure}[!ht]
    \footnotesize
    \centering
    \begin{subfigure}[t]{0.32\textwidth}
		\includegraphics[width=\textwidth]{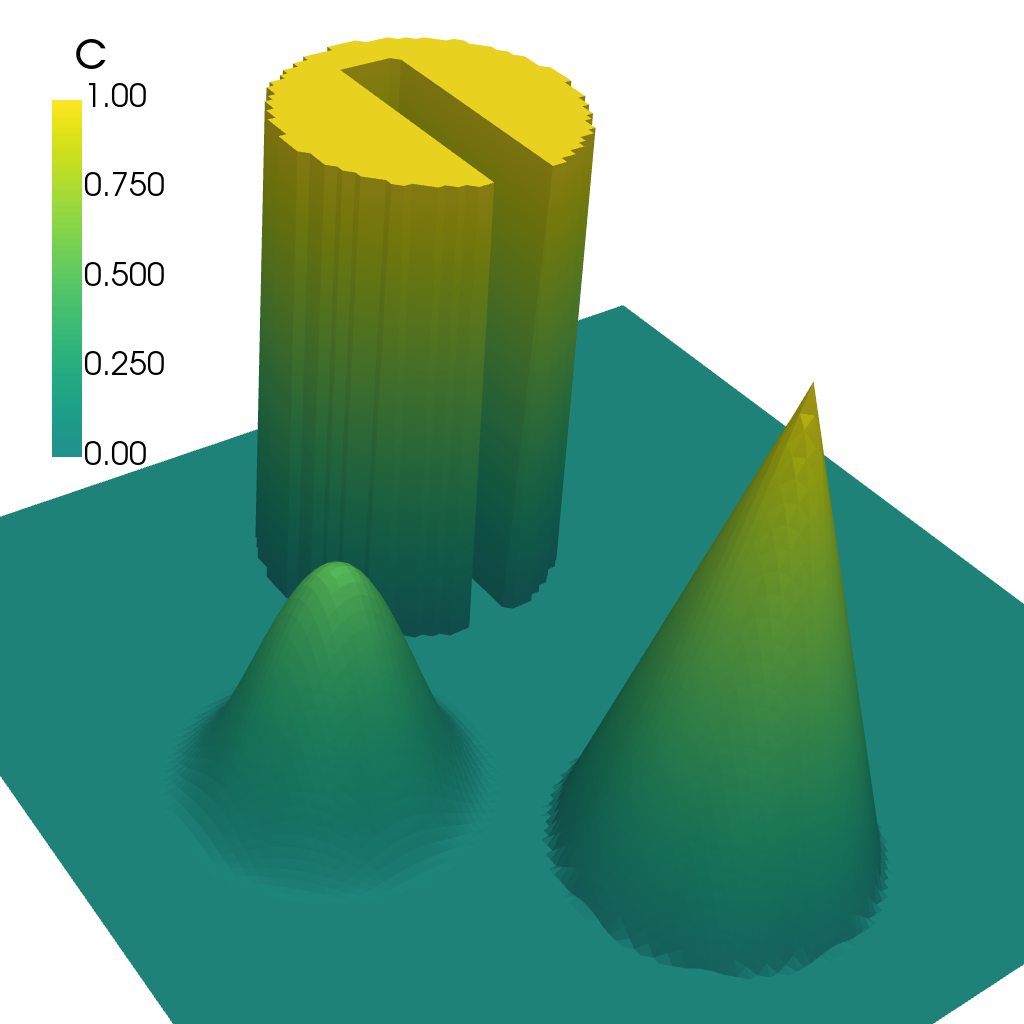}
        \caption{interpolated solution}
        \label{fig:benchmark-01-2d-initial}
	\end{subfigure}
    \hfill
    \begin{subfigure}[t]{0.32\textwidth}
		\includegraphics[width=\textwidth]{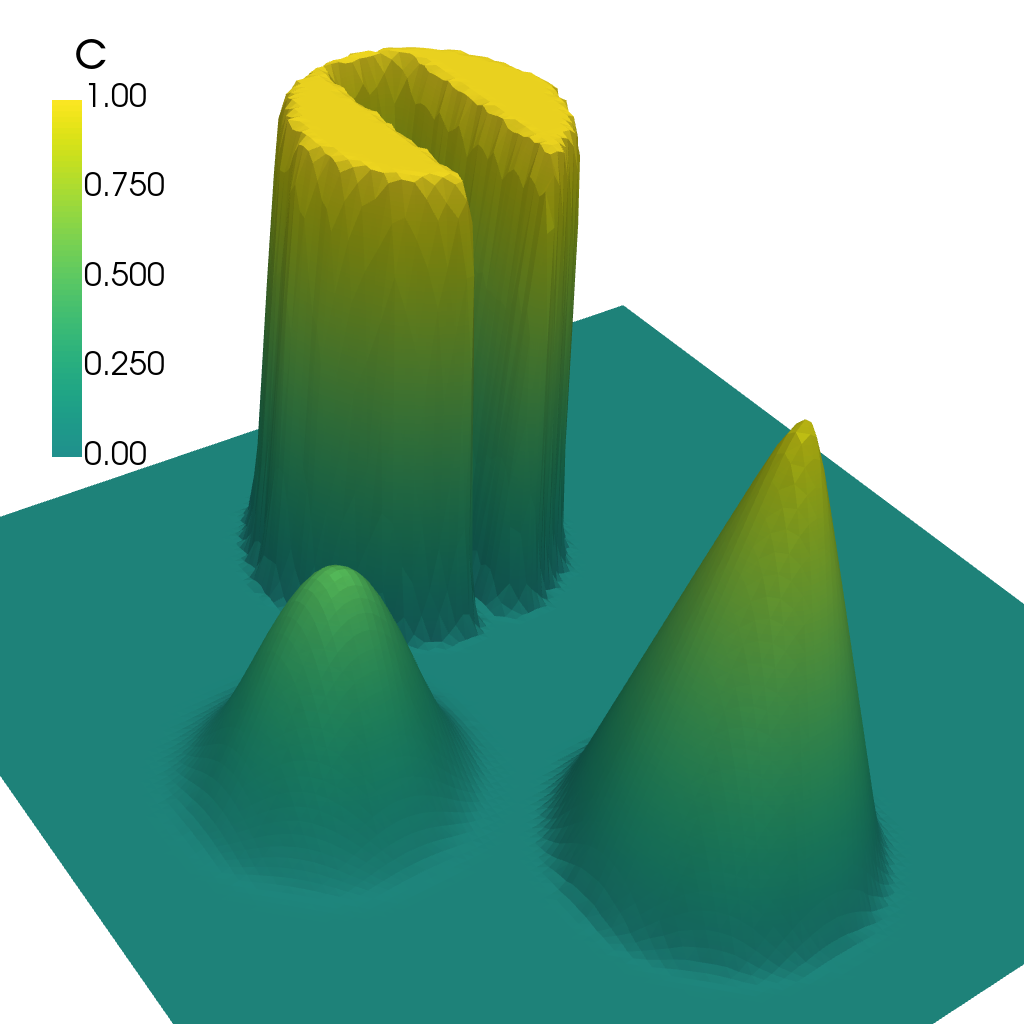}
        \caption{computed solution, $b = 1000$}
	\end{subfigure}
    \hfill
    \begin{subfigure}[t]{0.32\textwidth}
		\includegraphics[width=\textwidth]{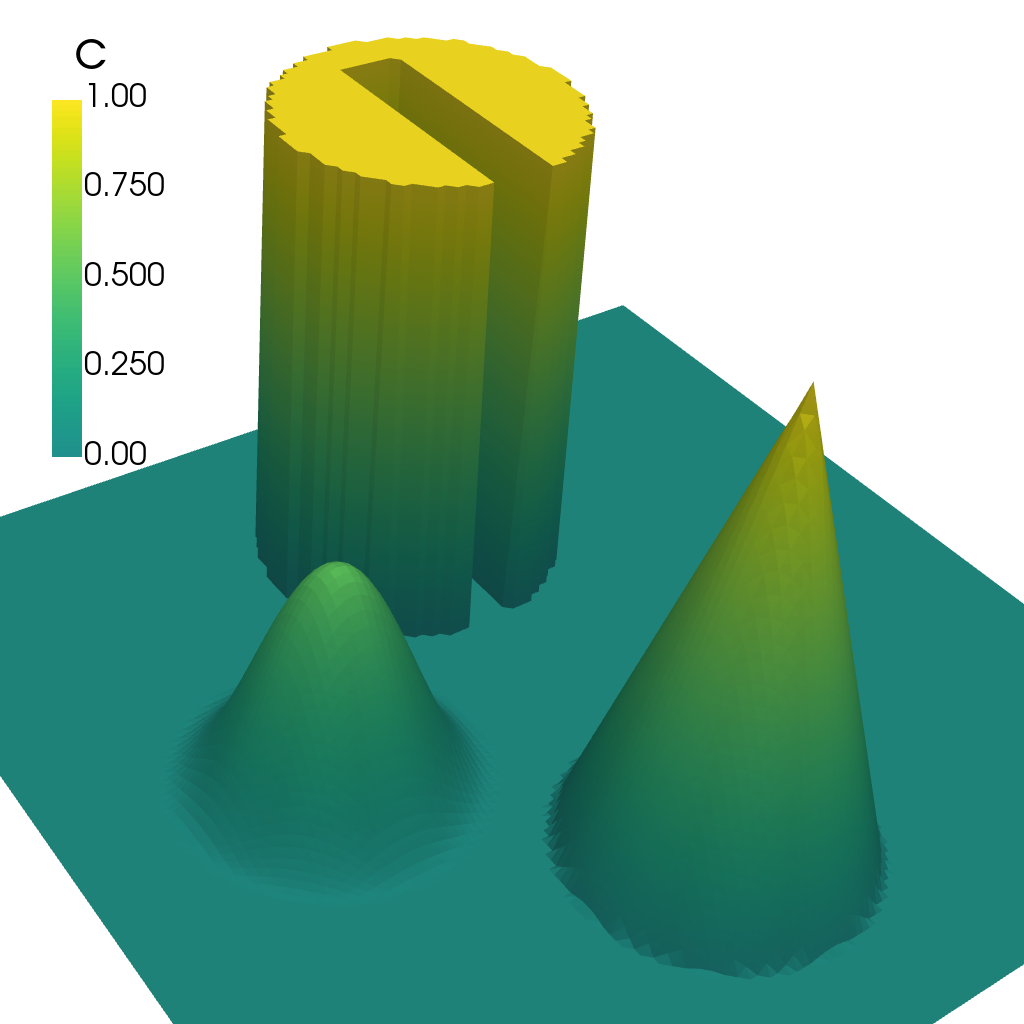}
        \caption{computed solution, $b = \infty$}
	\end{subfigure}
    \caption{Interpolated and computed solutions of the body rotation problem for the setup as in \cref{tab:benchmark-01-2d} with \Pone
    discretization for different look-back distances $b$. The plots show the influence of the look-back distance on the quality of the computed solution.}
    \label{fig:benchmark-01-2d}
\end{figure}

Let $\Omega = (0, 1)^2$ be the domain where the initial temperature is imposed by three bodies as shown in \cref{fig:benchmark-01-2d-initial}.
All bodies are defined on circles with radius $r_0 = 0.15$, the initial condition is zero
outside of these circles. 
We define $\mathbf{x} = (x_1, x_2)$, $\bar{\mathbf{x}} = (\bar{x}_1, \bar{x}_2)$,
$r(\mathbf{x}) := \norm{\mathbf{x} - \bar{\mathbf{x}}}_2 / r_0$,
and the initial condition $c_0 = c_0^\text{slotted} + c_0^\text{cone} + c_0^\text{hill}$ by
\begin{align}
    c_0^\text{slotted}(\mathbf{x}) &= 
    \begin{cases}
        1 \quad & \text{if } r(\mathbf{x}) \leq 1,\ |\mathbf{x} - \bar{\mathbf{x}}| \geq 0.025,\ y \geq 0.85 \\
        0 \quad &\text{otherwise}
    \end{cases} & \bar{\mathbf{x}} = (0.5, 0.75),\\
    c_0^\text{cone}(\mathbf{x}) &=
    \begin{cases}
        1 - r(\mathbf{x}) \quad & \text{if } r(\mathbf{x}) \leq 1 \\
        0           \quad & \text{otherwise}
    \end{cases} & \bar{\mathbf{x}} = (0.5, 0.25),\\
    c_0^\text{hill}(\mathbf{x}) &=
    \begin{cases}
        \frac{1}{4} ( 1 + \cos(\pi r(\mathbf{x})) ) \quad & \text{if } r(\mathbf{x}) \leq 1 \\
        0                                     \quad & \text{otherwise}
     \end{cases} & \bar{\mathbf{x}} = (0.25, 0.5).
\end{align}

The bodies are rotating counter-clockwise along the constant velocity field $\mathbf{u} = (0.5 - x_2, x_1 - 0.5)^\top$.
Since we consider pure advection ($\kappa=0$, $q=0$), at $t = 2\pi$, the bodies have finished a full revolution and the resulting
temperature field should be equal to the initial condition. The time step size $\tau$ is constant.

In \cref{tab:benchmark-01-2d}, the different versions of the \gls*{mmoc} are compared to the linear and non-linear 
\gls*{fct} methods that performed best in \cite{John:2008:CMAME}. We observe the strong influence of the 
look-back distance $b$ on the solution, as visualized in the plots of the computed solutions in 
\cref{fig:benchmark-01-2d}.

\begin{table}[!ht]
\centering
\footnotesize
\caption{Comparison of different parameterizations of the \gls*{mmoc} to the best performing methods from the study in \cite{John:2008:CMAME}.
         The FEM-FCT employs \Pone finite elements and a grid spacing of $h = 1/128$. We run our \gls*{mmoc} implementation both, with
         \Pone and \Ptwo elements and grid spacings of $h = 1/128$ and $h = 1/64$ respectively. The mesh size $h$ refers
         to quadratic elements, which are divided into two triangles each.
         Therefore, all grids result in 16641 \glspl*{dof}, including boundary. The time-step size is $\tau \approx \num{1e-03}$ for all
         settings (which corresponds to 6283 time steps for one revolution).}
\label{tab:benchmark-01-2d}
\resizebox{\columnwidth}{!}{%
\begin{tabular}{lrrr}
  \toprule
  method                               &   $\norm{e_h}_{\mathcal{H}^0}$ &   $\text{var}(2\pi)$ &   $\Delta m(2\pi)$ \\
  \midrule
  MMOC \Pone ($b = 1$)                   & \num{1.74e-01} & \num{0.5913} & \num{-4.73e-02} \\
  MMOC \Pone ($b = 10$)                  & \num{1.65e-01} & \num{0.6296} & \num{ 4.76e-03} \\
  MMOC \Pone ($b = 100$)                 & \num{8.60e-02} & \num{0.9847} & \num{ 2.67e-04} \\
  MMOC \Pone ($b = 1000$)                & \num{3.85e-02} & \num{1.0000} & \num{-5.52e-04} \\
  MMOC \Pone ($b = \infty$)              & \num{1.38e-13} & \num{1.0000} & \num{ 2.22e-16} \\
  FEM-FCT n.-l. \cite{John:2008:CMAME} & \num{1.44e-01} & \num{1.0010} & no data \\
  FEM-FCT \cite{John:2008:CMAME}       & \num{1.92e-01} & \num{1.0069} & no data \\
  \bottomrule
\end{tabular}
\hspace{3pt}
\begin{tabular}{lrrr}
  \toprule
  method                 &   $\norm{e_h}_{\mathcal{H}^0}$ &   $\text{var}(2\pi)$ &   $\Delta m(2\pi)$ \\
  \midrule
  MMOC \Ptwo ($b = 1$)      & \num{1.09e-01} & \num{1.2773} & \num{-2.19e-02} \\
  MMOC \Ptwo ($b = 10$)     & \num{9.71e-02} & \num{1.2943} & \num{-1.39e-02} \\
  MMOC \Ptwo ($b = 100$)    & \num{5.29e-02} & \num{1.3049} & \num{-5.56e-03} \\
  MMOC \Ptwo ($b = 1000$)   & \num{3.03e-02} & \num{1.3185} & \num{-1.09e-03} \\
  MMOC \Ptwo ($b = \infty$) & \num{1.68e-13} & \num{1.0000} & \num{-6.79e-14} \\
  \bottomrule
\end{tabular}
}
\end{table}

Optimal results are achieved with infinite look-back distance ($b=\infty$). This suggests that 
the interpolation between Eulerian and Lagrangian representation is the primary source of errors, and 
energy difference. We note, that in a massively parallel setting, occasional interpolation to the Eulerian domain may be desired to
reduce the communication overhead during the temperature evaluation (step (iii) in \cref{sec:particle-tracing}).

\begin{remark}[Choice of space-discretization, oscillations]\label{remark:space-discretization}
    The amount of spurious oscillations denoted by $\text{var}(2\pi)$ in \cref{tab:benchmark-01-2d} with
    \Ptwo elements and $b=\infty$ is partly misleading. While there are no oscillations at time $t = 2\pi$, some oscillations
    appear at the discontinuity around the slotted cylinder for $0 < t < 2\pi$. 
    For the linear (\Pone) space discretization, there are no oscillations over the entire time interval.

    Typically, continuous Lagrange finite elements of higher order tend to produce over- and
    undershoots at discontinuities. However, we note, that this is owed to the 
    space-discretization and not to the presented time-discretization, \ie the \gls*{mmoc}. An advantage
    of the \gls*{mmoc} is, that it can be applied to any space-discretization as long as the solution can be locally 
    evaluated. In presence of discontinuities in the solution, discontinuous Galerkin 
    space-discretizations could be considered in combination with the \gls*{mmoc}.
\end{remark}

To demonstrate, the contribution of the discontinuity around the slotted cylinder to the error, 
we show in \cref{tab:benchmark-01-2d-gaussian-only} results for the smooth 
initial condition and solution $c_0 = c_0^{\text{hill}}$. In this run,
the error and energy discrepancy is much smaller than for the results with a discontinuous solution
in \cref{tab:benchmark-01-2d}, especially, for $b < \infty$.

\begin{figure}[!ht]
    \centering
    \begin{minipage}{0.49\textwidth}
        \centering
        \resizebox{\textwidth}{!}{%
        \begin{tabular}{crrcr}
          \toprule
          $\tau$ &   $b$ &
          \multicolumn{1}{c}{$\norm{e_h}_{\mathcal{H}^0}$} & 
          $\text{var}(2\pi)$ &
          \multicolumn{1}{c}{$\Delta m(2\pi)$} \\
          \midrule
          \num{1.01e-01} &        1 & \num{3.36e-04} & \num{0.5019} & \num{-6.63e-05} \\
          \num{1.01e-01} &       10 & \num{5.32e-05} & \num{0.5008} & \num{ 3.83e-05} \\
          \num{1.01e-01} & $\infty$ & \num{9.62e-07} & \num{0.5000} & \num{ 9.35e-07} \\
          \midrule
          \num{1.00e-02} &        1 & \num{4.33e-03} & \num{0.5054} & \num{-4.46e-06} \\
          \num{1.00e-02} &       10 & \num{3.43e-04} & \num{0.5020} & \num{-5.33e-05} \\
          \num{1.00e-02} &      100 & \num{4.87e-05} & \num{0.5008} & \num{ 2.28e-05} \\
          \num{1.00e-02} & $\infty$ & \num{9.13e-11} & \num{0.5000} & \num{ 9.12e-12} \\
          \midrule
          \num{1.00e-03} &        1 & \num{6.33e-03} & \num{0.5066} & \num{-2.05e-05} \\
          \num{1.00e-03} &       10 & \num{4.33e-03} & \num{0.5054} & \num{-4.38e-06} \\
          \num{1.00e-03} &      100 & \num{3.40e-04} & \num{0.5020} & \num{-4.80e-05} \\
          \num{1.00e-03} &     1000 & \num{4.96e-05} & \num{0.5009} & \num{ 3.27e-05} \\
          \num{1.00e-03} & $\infty$ & \num{9.87e-15} & \num{0.5000} & \num{ 4.88e-15} \\
          \bottomrule
        \end{tabular}
      }
      \captionof{table}{Results for the circular advection benchmark with $c_0 = c_0^{\text{hill}}$ and a \Ptwo space discretization.}
      \label{tab:benchmark-01-2d-gaussian-only}
    \end{minipage}\hfill
    \begin{minipage}{0.49\textwidth}
        \centering
        \resizebox*{\textwidth}{!}{
            \input{figures/bm_01_rotation_time_step_conv.pgf}}
          \caption{Time-step size study for the circular advection benchmark with $c_0 = c_0^\text{slotted} + c_0^\text{cone} + c_0^\text{hill}$ and $b=\infty$.}
          \label{fig:benchmark-01-time-convergence}
    \end{minipage}
\end{figure}

\Cref{tab:benchmark-01-2d-gaussian-only} additionally lists results of simulations with time-steps that
are increased by a factor of 10 and 100.
The measured errors demonstrate that the Lagrangian approach yields promising stability and accuracy also for
comparatively large time-steps.

For $b < \infty$ the resulting errors are mostly caused by the interpolation between 
the Eulerian and the Lagrangian representation. 
If we compare runs where $\tau \cdot b = \text{const}$, we obtain almost identical errors
(\eg $\tau = \num{1.01e-01}$ and $b = 1$ compared to $\tau = \num{1.00e-02}$ and $b = 10$
in \cref{tab:benchmark-01-2d-gaussian-only}).
In those runs, the number of time-steps in which the solution is interpolated is equal.
Despite a significant time-step size reduction, the interpolation error dominates.
In the case of $b = \infty$ no temperature interpolation is performed throughout the simulation.
Therefore the increased accuracy of the \gls*{rk} integrator directly affects the
error in the solution when the time-step size is reduced.

In \cref{fig:benchmark-01-time-convergence} we plot the $\mathcal{H}^0$ error of the solution of the original benchmark problem 
($c_0 = c_0^\text{slotted} + c_0^\text{cone} + c_0^\text{hill}$, \ie with discontinuous solution) discretized with 
\Pone finite elements for different time-step sizes and $b=\infty$. For the largest time-step size in \cref{fig:benchmark-01-time-convergence}
($\tau \approx .065$) and a maximum absolute velocity of $\approx .7$ this results in a \gls*{cfl} number of
roughly $3$.

\subsection{Swirling advection}

Next we move to a three-dimensional setting with a time-dependent velocity field. The benchmark is taken from \cite{LeVeque:1996:SINUM}.
Let $\Omega = (0, 1)^3$ and $t \in [0, T]$ with $T = 1.5$. The the initial condition $c_0$ and the velocity 
field $\mathbf{u}(\mathbf{x}, t)$ are defined by
\begin{align}
    c_0(\mathbf{x}) := 
    \begin{cases}
        1 \quad \text{if } x_1 < 0.5 \\
        0 \quad \text{otherwise}
    \end{cases},
    \quad
    \mathbf{u}(\mathbf{x}, t) := 
    \begin{pmatrix}
        2 \sin^2(\pi x_1) \sin(2 \pi x_2) \sin(2 \pi x_3) g(t) \\
        - \sin(2 \pi x_1) \sin^2(\pi x_2) \sin(2 \pi x_3) g(t) \\
        - \sin(2 \pi x_1) \sin(2 \pi x_2) \sin^2(\pi x_3) g(t)
    \end{pmatrix},
\end{align}
with $g(t) := \cos(\pi t / T)$. 
The temperature field undergoes a deformation which reverses at $t = T/2$ and should 
return to the initial solution at $t = T$. Again, we consider pure advection ($\kappa=0$, $q=0$).

\begin{figure}[!ht]
    \footnotesize
    \centering
    \begin{subfigure}[t]{0.32\textwidth}
		\includegraphics[width=\textwidth]{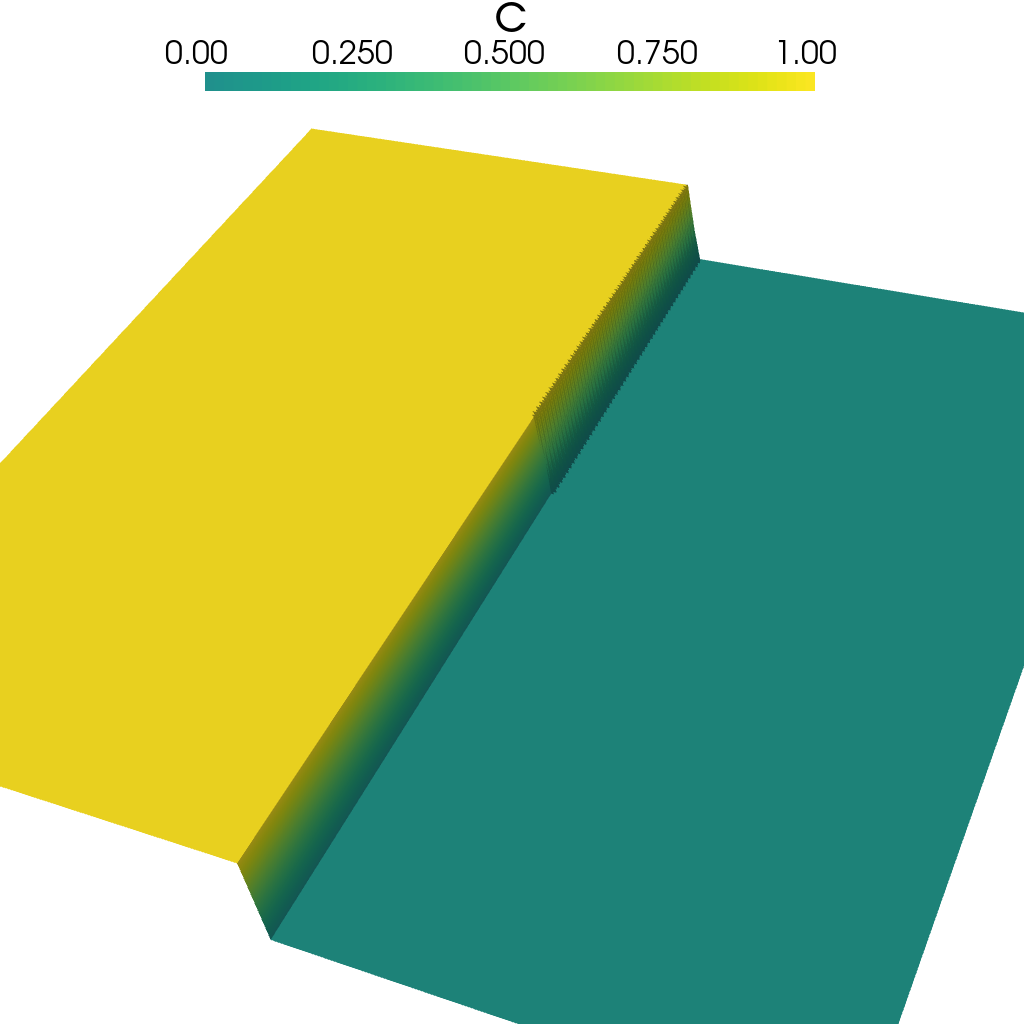}
        \caption{interpolated initial condition}
	\end{subfigure}
    \hfill
    \begin{subfigure}[t]{0.32\textwidth}
		\includegraphics[width=\textwidth]{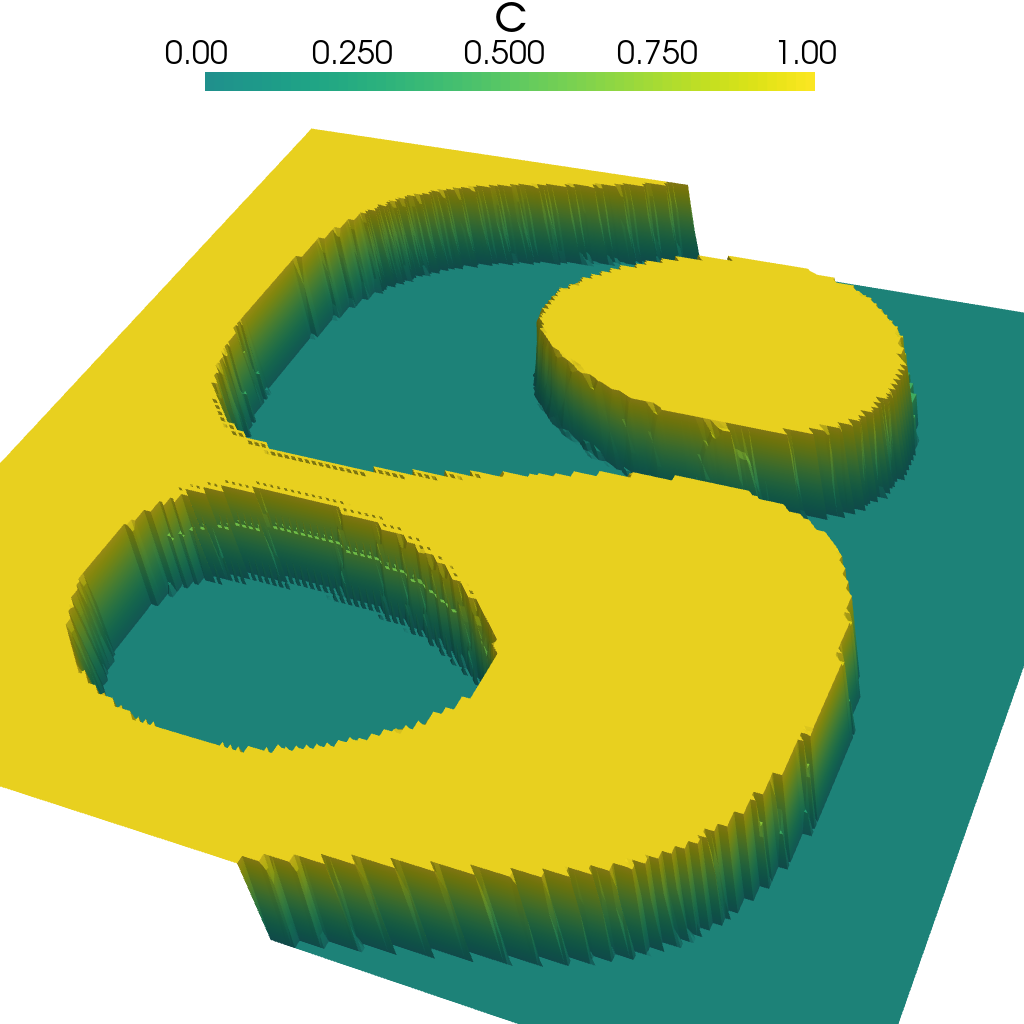}
        \caption{$t = T/2$ (computed solution)}
	\end{subfigure}
    \hfill
    \begin{subfigure}[t]{0.32\textwidth}
		\includegraphics[width=\textwidth]{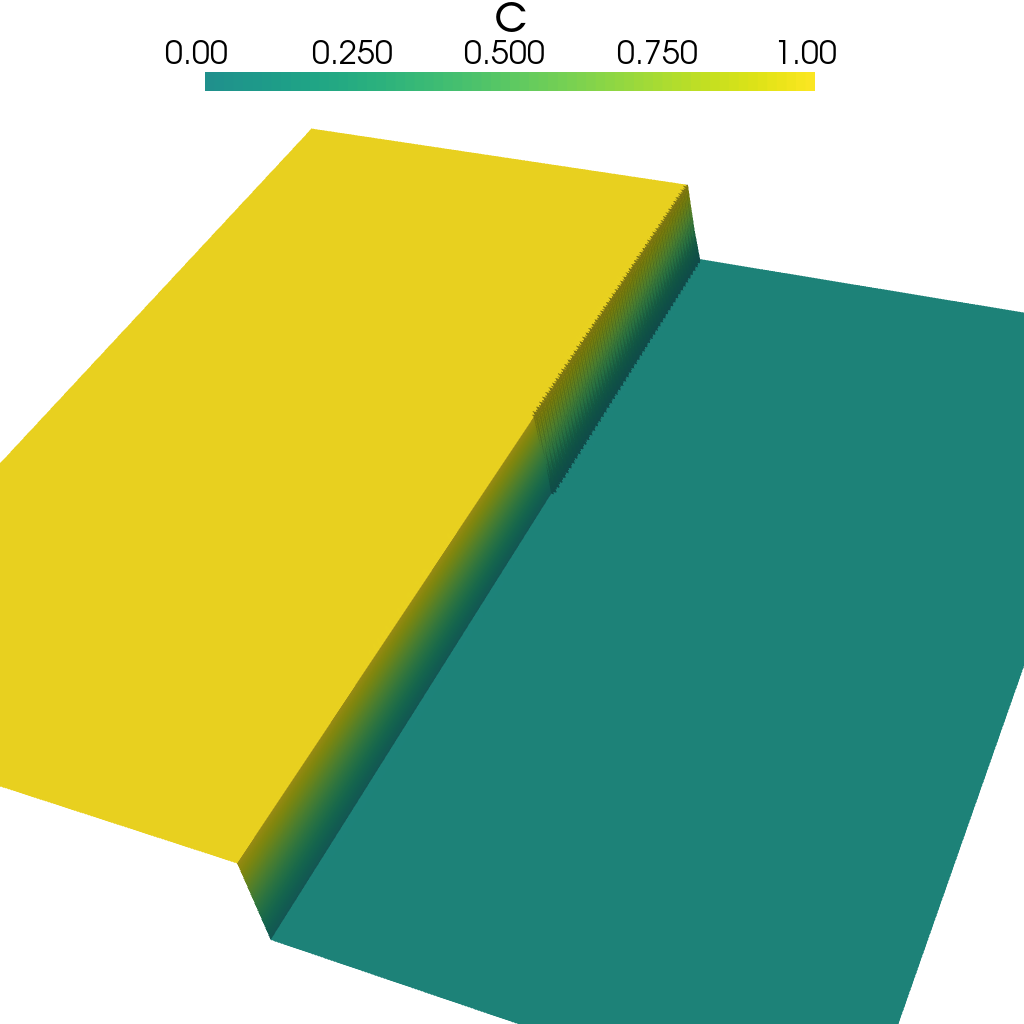}
        \caption{$t = T$ (computed solution)}
	\end{subfigure}
    \caption{Elevated slice of the computed solution at $x_3 = \num{0.425}$, as in \cite[figure 11.2]{LeVeque:1996:SINUM}.
    Parameters: \Pone discretization, $b=\infty$, $h_\text{min}=\num{7.81e-03}$ (refinement level 7),
    $\tau = \num{5.00e-02}$. The discontinuities are preserved without any oscillations or numerical diffusion.}
    \label{fig:benchmark-02}
\end{figure}

The results for the \gls*{mmoc} with $b=\infty$ at $T=1.5$ are listed in \cref{tab:benchmark-02}.
The $\mathcal{H}^0$-errors are small and no spurious oscillations are detected
for all chosen time step and grid sizes.
\begin{table}[!ht]
\centering
\footnotesize
\resizebox{0.7\textwidth}{!}{%
\begin{tabular}{rccccr}
    \toprule
    \multicolumn{1}{c}{\glspl*{dof}} & $h_\text{min}$ &   $\tau$ &   $\norm{e_h}_{\mathcal{H}^0}$ &   $\text{var}(1.5)$ &
    \multicolumn{1}{c}{$\Delta m(1.5)$} \\
    \midrule
    \numINT{  35937} & \num{3.12e-02} & \num{1.00e-01} & \num{8.67e-04} & \num{1.0000} & \num{-9.75e-05} \\
    \numINT{  35937} & \num{3.12e-02} & \num{5.00e-02} & \num{5.48e-05} & \num{1.0000} & \num{-2.37e-06} \\
    \numINT{  35937} & \num{3.12e-02} & \num{2.50e-02} & \num{5.11e-06} & \num{1.0000} & \num{-1.37e-08} \\
    \numINT{2146689} & \num{7.81e-03} & \num{1.00e-01} & \num{1.26e-03} & \num{1.0000} & \num{-2.77e-05} \\
    \numINT{2146689} & \num{7.81e-03} & \num{5.00e-02} & \num{5.01e-05} & \num{1.0000} & \num{-1.11e-06} \\
    \numINT{2146689} & \num{7.81e-03} & \num{2.50e-02} & \num{2.57e-06} & \num{1.0000} & \num{-2.18e-08} \\
    \bottomrule
\end{tabular}}
\caption{Results for application of the \gls*{mmoc} with $b=\infty$ and \Pone finite elements
in space to the swirling flow benchmark in 3D.}
\label{tab:benchmark-02}
\end{table}
\Cref{fig:benchmark-02} shows the computed solution at $x_3 = \num{0.425}$ for $t=T/2$, and $t=T$ with
$h_\text{min}=\num{7.81e-03}$ (refinement level 7), and $\tau = \num{5.00e-02}$. 
At $t=T$, the initial temperature field is restored without visible artifacts or numerical diffusion.
The slice is chosen to coincide with the slice shown in \cite[figure 11.2]{LeVeque:1996:SINUM}.

\subsection{Advection-diffusion on blended geometry}\label{sec:benchmark-advection-diffusion}

Finally we apply the \gls*{mmoc} to a problem with a diffusion coefficient $\kappa > 0$,
without internal heating ($q=0$), on a blended geometry. Since $\kappa > 0$, we need to solve the linear
system \cref{eq:linear-system} in each time-step. Thus the solution must be interpolated to the Eulerian grid
in each time-step and we must limit the look-back distance to $b = 1$.

The physical domain $\Omega_\text{phy}$ is an annulus defined by
$\Omega = \{ \mathbf{x} \in \mathbb{R}^2 : r_\text{min} \leq \norm{\mathbf{x}}_2 \leq r_\text{max} \}$
with $(r_\text{min}, r_\text{max}) = (0.5, 1.5)$. 
The computational domain $\Omega_\text{comp}$ approximates the annulus with a coarse 
triangular mesh that is uniformly refined and projected onto $\Omega_\text{phy}$, see 
\cref{fig:annulus-domain}.

The benchmark is inspired by the unsteady advection-diffusion benchmark in
\cite{Kuzmin:2004:CMAME}. A circular velocity field $\mathbf{u}(\mathbf{x}) = (-x_2, x_1)$
transports a gradually smeared Gaussian hill around the annulus.
The time-dependent position of the hill is given by an initial position 
$(\bar{x}_1, \bar{x}_2) = (0, 1)$ and $\hat{\mathbf{x}}(t) = (\hat{x}_1(t), \hat{x}_2(t))$ with
$\hat{x}_1(t) = \bar{x}_1 \cos(t) - \bar{x}_2 \sin(t)$, and $\hat{x}_2(t) = -\bar{x}_1 \sin(t) + \bar{x}_2 \cos(t)$.
The analytical solution $c$ is defined by
\begin{align}
    c(\mathbf{x}, t) := \frac{1}{4 \pi t \kappa} \exp\left(-\frac{r(\mathbf{x}, t)^2}{4t\kappa}\right),
    \quad \kappa > 0,\ t > 0\enspace,
\end{align}
where $r(\mathbf{x}, t) := \norm{\mathbf{x} - \hat{\mathbf{x}}}_2$.
At time $t = 0$, $c$ becomes a Dirac delta function, which is why we start the
simulation at a later time. To better compare the proposed method for different
$\kappa$, we parameterize the initial time $t_0$ via $t_0(\kappa) = \frac{2 \pi \times 10^{-3}}{\kappa}$.
The moving hill has therefore a different initial position when applying a
different diffusion coefficient, but the initial shape is identical for 
all choices of $\kappa$. The computed solution is evaluated after a full revolution,
at time $T = t_0(\kappa) + 2\pi$. We employ a \Ptwo space-discretization in all runs.
Results after one revolution are listed in \cref{tab:benchmark-04}.
\begin{table}[!ht]
\centering
\footnotesize
\resizebox{\columnwidth}{!}{%
\begin{tabular}{rr|rr|rr|rr}
  \multicolumn{2}{c}{} & \multicolumn{2}{c}{$\kappa = \num{1e-3}$} & \multicolumn{2}{c}{$\kappa = \num{1e-5}$} & \multicolumn{2}{c}{$\kappa = \num{1e-7}$} \\
  \midrule
  \glspl*{dof} &   $h_\text{min}$ &   $\norm{e^h}_{\mathcal{H}^0}$ &   $E_\text{peak}(2\pi)$ &   $\norm{e^h}_{\mathcal{H}^0}$ &   $E_\text{peak}(2\pi)$ &   $\norm{e^h}_{\mathcal{H}^0}$ &   $E_\text{peak}(2\pi)$ \\
  \midrule
  \num{ 12480} & \num{3.12e-02} & \num{1.48e-02} & \num{-1.30e-03} & \num{8.32e-02} & \num{-2.02e-02} & \num{8.51e-02} & \num{-1.98e-02} \\
  \num{ 49536} & \num{1.56e-02} & \num{3.86e-03} & \num{ 3.45e-03} & \num{7.38e-03} & \num{-1.90e-03} & \num{7.84e-03} & \num{-1.56e-03} \\
  \num{197376} & \num{7.81e-03} & \num{4.32e-03} & \num{ 3.98e-03} & \num{6.30e-04} & \num{-1.25e-04} & \num{6.92e-04} & \num{-1.01e-04} \\
 \bottomrule
\end{tabular}}
\caption{Results after one revolution of the Gaussian hill, with time-step size $\tau \approx 1e-1$,
implicit Euler time-integration for the parabolic part ($\Theta = 1$ in \cref{eq:finite-dimensional-galerkin-approximation}),
three different diffusion coefficients $\kappa$, and different refinement levels.}
\label{tab:benchmark-04}
\end{table}
For a comparably large time step size (\gls*{cfl} between 4 and 20) and varying diffusivity, we observe
satisfying results for sufficiently small mesh sizes. 

\section{Parallel performance}\label{sec:performance}

Scalability of the HHG data structures in \gls*{hyteg}, and of the particle dynamics framework
\gls*{mesapd} has separately been demonstrated on some of the worlds largest supercomputers,
\cite{Kohl:2020:arXiv,Eibl:2018:PARCO}. It remains to assess the parallel performance of our
\gls*{mmoc} implementation, in which both software architectures are coupled.

For the scalability benchmark, we set up an elongated, three-dimensional cuboid domain, where a smooth 
initial temperature field is initialized, and transported along a constant 
velocity field, resembling flow through a pipe. This setup allows for straightforward parameterization
of domain size, and number of coarse grid primitives.

We employ \Ptwo finite elements for the space-discretization and perform a single time-step,
consisting of particle creation, particle integration, and temperature evaluation, including synchronization
(steps (i)--(iii) in \cref{sec:particle-tracing}).

All runs in this section were performed on \supermucng, ranked 15\textsuperscript{th} in the 
Top500\footnote{\url{https://www.top500.org/}} list (Nov 2020). The system is composed of \numINT{6336} so-called
thin-nodes, \numINT{3072} of which we had access to at the time of writing. Two 
Intel\textsuperscript{\textregistered} Skylake Xeon\textsuperscript{\textregistered} Platinum 8174 CPUs
are installed on each node, which sums up to 48 cores per node and \numINT{147456} cores in total on the
accessible \numINT{3072} nodes. Per node $96$GB of main memory are available.

\paragraph{Strong scaling}

\begin{figure}[!ht]
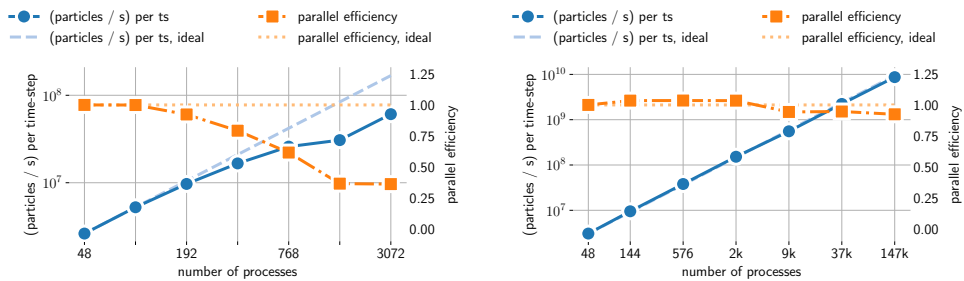

  \footnotesize
  \centering
  \begin{subfigure}[t]{0.49\textwidth}
    \resizebox*{\textwidth}{!}{%
      \input{figures/scaling_strong.pgf}
    }
    \caption{strong scaling, $\approx 1.7\times10^7$ temperature \glspl*{dof} (\ie particles) in total}
    \label{fig:scaling-strong}
  \end{subfigure}
  \hfill
  \begin{subfigure}[t]{0.49\textwidth}
    \resizebox*{\textwidth}{!}{%
      \input{figures/scaling_weak.pgf}
    }
    \caption{weak scaling, $\approx 3.55 \times 10^{5}$ temperature \glspl*{dof} (\ie particles) per process}
    \label{fig:scaling-weak}
  \end{subfigure}
  \caption{Weak and strong scaling results (ts = time-step)}
  \label{fig:scaling}
\end{figure}

We conduct a strong scaling experiment on a grid that consists of \numINT{3072} tetrahedral coarse grid elements,
each of which is refined 4 times, resulting in 
$\approx \num{1.7e7}$ \glspl*{dof} in total.
Leaving the grid fixed, we increase the number of processes, so that the number of particles per process decreases.
As a baseline for the parallel performance we consider a single-node run.
We plot the parallel performance and the number of updated particles per second in \cref{fig:scaling-strong}.
For the largest setting with 64 nodes (\numINT{3072} processes) we obtain a parallel efficiency of roughly
\percent{36} for one macro-cell and $\approx \numINT{5500}$ particles per process.

\paragraph{Weak scaling}

Additionally, we perform a weak-scaling experiment, where the number of \glspl*{dof} per process is kept constant.
In this setting, each process is assigned a single tetrahedral macro-cell and we refine the initial grid 5 times.
This results in about \num{3.55e5} \glspl*{dof} per process. Starting from a single node,
again used as baseline for parallel efficiency,
we scale up to the available \numINT{147456} processes of \supermucng.
In the largest scenario, this amounts to more than \num{5.2e10} \glspl*{dof} in total for the discretization
of the solution of the advection-diffusion equation \cref{eq:advection-diffusion-pde}. 
All runs maintain an excellent parallel efficiency of more than \percent{92}.

Overall, we observe a run time per time step of about 5 or less seconds in all tested scenarios, and less
than a second in the strong-scaling limit. We note, however, that it is not sufficient to consider the number of 
updated \glspl*{dof} per second alone as a measure to quantify the efficiency of the method. 
The results of \cref{sec:benchmarks} show that the stability and accuracy of the \gls*{mmoc} allows for 
large time-steps even in the strongly advection-dominated problems. 
This may be an advantage in coupled convection simulations as they appear in Earth mantle convection, where
the majority of the run time is spent for the solution of the Stokes system \cite{Kronbichler:2012:GJI,Gmeiner:2015:SISC}. 
Given sufficiently accurate coupling schemes, the \gls*{mmoc} does not only allow for a reduction of simulation 
time by itself, but also permits to advance faster in time, due to less restrictive \gls*{cfl} limitations.

\section{Coupled flow}\label{sec:coupled-flow}

Finally, we apply our scheme to a buoyancy-driven, coupled flow problem.
Due to the
negligible Reynolds number
in mantle convection models, the Stokes equations are used to model the creeping flow
of the medium. We consider the incompressible formulation for the Boussinesq approximation \cite{Ricard:2007:Treatise}
\begin{equation}\label{eq:stokes-pde}
  \begin{aligned}
	- \nabla \cdot \sigma = \mathbf{F}(c)&, \quad
	\nabla \cdot \mathbf{u} = 0 \\
  \sigma(\mathbf{u}, p) = 2 \mu \epsilon(\mathbf{u}) - pI&, \quad
	\epsilon(\mathbf{u}) = \frac{1}{2}\left(\nabla \mathbf{u} + (\nabla \mathbf{u})^\top\right).
  \end{aligned}
\end{equation}
where $\mu$ is a viscosity field, $p$ the pressure, $\mathbf{F}(c) := \text{Ra}\,c\,\mathbf{g}$ a temperature dependent forcing term,
$\text{Ra}$ the Rayleigh number, $\mathbf{g}$ the normalized gravitation, and $\sigma$ 
the Cauchy stress tensor associated with an incompressible, highly viscous Newtonian fluid.

The \glspl*{pde} \cref{eq:advection-diffusion-pde} and \cref{eq:stokes-pde} are coupled through both,
the velocity which is the solution of the Stokes system and drives the advection of the temperature,
and the temperature which enters the Stokes equation through the forcing term.
Different from the benchmarks in \cref{sec:benchmarks}, the convectivity is in the following setups steered
solely through the Rayleigh number $\text{Ra}$, \ie we set $\kappa = 1$ in \cref{eq:advection-diffusion-pde}. When $\text{Ra}$ is large,
so is the \gls*{rhs} of \cref{eq:stokes-pde} and the velocity that enters \cref{eq:advection-diffusion-pde}
has a large magnitude, resulting in advection-dominated transport. Note that since $\kappa > 0$, the look-back distance
is set to $b = 1$ in the following benchmarks.

The advection-diffusion equation \cref{eq:advection-diffusion-pde} is constrained by the Stokes equation \cref{eq:stokes-pde} at all times,
and a non-linear system must be solved at each time-step, which is at least a computationally expensive challenge for large-scale simulations.
In practice, \cref{eq:advection-diffusion-pde} is thus usually \emph{decoupled} from the 
constraints to the velocity $\mathbf{u}$, so that the systems can be solved in an alternating 
fashion \cite{Kronbichler:2012:GJI,Waluga:2016:JCP}. 

The solution of the Stokes system in each time-step dominates the computational cost of this scheme and
is therefore crucial to performance. We employ an efficient, monolithic matrix-free geometric multigrid solver as described
in \cite{Kohl:2020:arXiv}, and large \gls*{cfl}-numbers to reduce the number of required solves.
The Stokes system is discretized with a mixed \Ptwo-\Pone finite element approximation.
For more in-depth discussion of efficient matrix-free geometric multigrid solvers on \gls*{hhg},
we refer to \cite{Kohl:2020:arXiv,Bauer:2020:SPPEXA,Bauer:2017:ANM,Bauer:2019:JoCS,Gmeiner:2016:JoCS,Gmeiner:2015:SISC}.

In \cref{sec:pc}, we outline a predictor-corrector scheme (see \cref{alg:pc}) to approximate the solution of the non-linear, coupled system, 
and apply the method to two benchmark problems. 

\subsection{Strang-splitting}

We employ a tighter coupling of the advection- and diffusion-step via a Strang-splitting
approach \cite{Strang:1968:SINUM}. Instead of an alternating application of the advection and diffusion step,
the diffusion step is split, and the advection step is framed by two fractional diffusion steps with reduced time-step size,
giving a scheme with three stages. The algorithm is listed in \cref{alg:ad-splitting}.
\begin{algorithm}
    \footnotesize
    \begin{algorithmic}[1] 
		\Procedure{ADS}{$c_h^n, \mathbf{u}_h$}
			\State $\tau_n = \text{CFL}_\text{max} \cdot h_\text{min} / \max_{\mathbf{x}\in\Omega}|\mathbf{u}_h(\mathbf{x}, t_n)|$  
				\Comment{determine time-step size}
			\State solve \cref{eq:linear-system} with $\tau_n^* = \tau_n/2$ to advance from $c_h^{n}$ to $c_h^{n+(1/3)}$
				\Comment{diffusion}
			\State $\hat{\mathbf{x}} = \mathbf{X}(\mathbf{x}, t_{n+1}, t_{n})$ 
				\Comment{calculate departure points (see \cref{sec:implementation})}
			\State $c_h^{n+(2/3)}(\mathbf{x}) = c_h^{n+(1/3)}(\hat{\mathbf{x}})$
				\Comment{advection}
			\State solve \cref{eq:linear-system} with $\tau_n^* = \tau_n/2$ to advance from $c_h^{n+(2/3)}$ to $c_h^{n+1}$
				\Comment{diffusion}
			\State \textbf{return} $c_h^{n+1}$
        \EndProcedure
    \end{algorithmic}
    \caption{\footnotesize Time-stepping scheme, advection-diffusion, with Strang-splitting}
    \label{alg:ad-splitting}
\end{algorithm}

The splitting procedure noticeably increases the accuracy of the method in the benchmarks of this section, however,
we did not observe relevant differences when applying it to the advection-diffusion benchmark in \cref{sec:benchmark-advection-diffusion}.

\subsection{A predictor-corrector scheme}\label{sec:pc}

To resolve the non-linear coupling of the advection-diffusion equation \cref{eq:advection-diffusion-pde}
and the Stokes problem \cref{eq:stokes-pde}, we apply a predictor-corrector method \cite{vandenBerg:1993:GJI}, 
as outlined in \cref{alg:pc}.

\begin{algorithm}
    \footnotesize
    \begin{algorithmic}[1] 
		\Procedure{PC}{}
			\State solve \cref{eq:stokes-pde} for $\mathbf{u}_h^0$
				\Comment{initial velocity field}
			\For{$n \in \{0, 1, \dots\}$}
				\State $\tilde{\mathbf{u}}_h(\mathbf{x}, t) \gets \mathbf{u}_h^n(\mathbf{x})$
					\Comment{time-invariant velocity field at time-step $n$ for temp. predictor}
				\State $c_h^\text{pr} \gets$ \Call{ADS}{$c_h^n, \tilde{\mathbf{u}}_h$}
					\Comment{predict temperature}
				\State solve \cref{eq:stokes-pde} for $\mathbf{u}_h^\text{pr}$ with RHS $\mathbf{F}(c_h^{\text{pr}})$
					\Comment{predict velocity}
				\State $\tilde{\mathbf{u}}_h(\mathbf{x}, t) \gets \text{lerp}(\mathbf{u}_h^n, \mathbf{u}_h^\text{pr}, t)(\mathbf{x})$
					\Comment{linear interpolation in $t$ between $\mathbf{u}_h^n$ and $\mathbf{u}_h^\text{pr}$}
				\State $c_h^{n+1} \gets$ \Call{ADS}{$c_h^n, \tilde{\mathbf{u}}_h$}
					\Comment{correct temperature}
				\State solve \cref{eq:stokes-pde} for $\mathbf{u}_h^{n+1}$ with RHS $\mathbf{F}(c_h^{n+1})$
					\Comment{correct velocity}
			\EndFor
        \EndProcedure
    \end{algorithmic}
	\caption{\footnotesize Predictor-corrector scheme to couple \cref{eq:advection-diffusion-pde} and \cref{eq:stokes-pde}. 
	In each time-step both \glspl*{pde} are solved twice. For the advection-diffusion step, \cref{alg:ad-splitting}
	is employed.} 
    \label{alg:pc}
\end{algorithm}

For the temperature prediction step, we approximate the velocity field with the time-invariant state at $t = t_n$,
\ie the interpolation \cref{eq:linear-interpolation} yields $\mathbf{u}_h^n$ for all $t \in [t_n, t_{n+1}]$.
A prediction $\mathbf{u}_h^\text{pr}$ for the velocity is then computed using the predicted temperature field for the 
\gls*{rhs} force term of \cref{eq:stokes-pde}.
The correction step is then executed, employing the interpolation in \cref{eq:linear-interpolation} between
$\mathbf{u}_h^n$ at $t = t_n$, and $\mathbf{u}_h^\text{pr}$ at $t = t_{n+1}$. Finally a new velocity solution is computed
using the corrected temperature field.

\subsection{Time-dependent convection benchmark}\label{sec:blankenbach}

To verify our implementation, we consider a classical benchmark from Blankenbach et
al.~\cite{Blankenbach:1989:GJI} (case 3) that was also investigated e.g.~in
\cite{Vynnytska:2013:C+G}. The test considers time-dependent convection with
constant viscosity ($\mu = 1$ in \cref{eq:stokes-pde}) and internal heating
($q = 1$ in \cref{eq:advection-diffusion-pde}) in a two-dimensional,
rectangular domain $\Omega = [0, L] \times [0, H]$, $L = 1.5, H = 1$. The top, bottom,
and side boundaries are denoted as $\Gamma_t$, $\Gamma_b$, and $\Gamma_s$.
$\mathbf{n}$ and $\mathbf{t}$ represent the outward normal, and tangential vectors
respectively.
For the velocity free-slip conditions are prescribed at the vertical boundaries
($\mathbf{u} \cdot \mathbf{n} = \sigma \mathbf{n} \cdot \mathbf{t} = 0$ for
$\mathbf{x} \in \Gamma_s$), and no-slip conditions at the horizontal boundaries
($\mathbf{u} = 0$ for $\mathbf{x} \in \Gamma_t \cup \Gamma_b$). For the temperature
zero Dirichlet boundary conditions are prescribed at the top boundary
($c = 0$ for $\mathbf{x} \in \Gamma_t$), and Neumann boundaries otherwise
($\partial_\mathbf{n} c = 0$ for $\mathbf{x} \in \Gamma_b \cup \Gamma_s$).
We employ the initial condition 
$c_0(\mathbf{x}) = 0.5 \left(1-x_2^2\right) + 0.01 \cos(\pi x_1 / L) \sin(\pi x_2 / H)$
given in \cite{Vynnytska:2013:C+G}.

The benchmark solution is expected to exhibit a characteristic, periodic development of
downwelling plumes and is quantified via the local extrema of the root-mean-square 
velocity $\mathbf{u}_\text{rms}$ and the Nusselt number $\text{Nu}$, defined as
\begin{align}\label{eq:rms}
  \mathbf{u}_\text{rms} = \left( \frac{1}{|\Omega|} \int_\Omega \norm{\mathbf{u}}^2 dx \right)^{1/2}, \quad
  \text{Nu} = - \frac{\int_0^L \partial_{x_2} c(\mathbf{x}, x_2 = H) dx}{\int_0^L c(\mathbf{x}, x_2 = 0) dx}
  \enspace.
\end{align}
At low Rayleigh numbers, every plume shows the same behavior. With increasing $\text{Ra}$ the periodicity
is characterized by every $n$-th plume behaving identically, resulting in a $Pn$-cycle. In particular,
the benchmark suggests that the convective motion transitions from a $P2$- to a $P4$-cycle between
$\text{Ra} = \numINT{216000}$ and $\text{Ra} = \numINT{218000}$. We partition a Pn-cycle into $n$ time 
intervals, denoted as stages $S0, \dots, S(n-1)$.
Each stage of a cycle comprises a local maximum of $\mathbf{u}_\text{rms}$ and $\text{Nu}$,
followed by a local minimum.

We apply the time-stepping scheme \cref{alg:pc} on two meshes with sizes $24 \times 16$ and $48 \times 32$ squares 
(each divided into 2 triangles) and for two CFL-numbers (\num{0.5} and \num{1}) running the simulation
from $t = 0$ to $t=3$.
For $t \in [2.5, 3]$, the described repetitive cyclic motion of the plumes is observed.
The computed solution is compared to the reference values in \cite[table 9]{Blankenbach:1989:GJI}
for $\text{Ra} = \numINT{216000}$, and \cite[table 8a, Code Ha, $96 \times 64$]{Blankenbach:1989:GJI}
for $\text{Ra} = \numINT{218000}$. We selected the latter reference from the various codes compared in
\cite{Blankenbach:1989:GJI} as the presumably most accurate implementation, and note that no
analytical solution is known.

The relative errors (compared to the reference) of the minima and maxima of $\text{Nu}$ and $\mathbf{u}_\text{rms}$ are calculated.
For $\text{Ra} = \numINT{216000}$, all extrema coincide with the reference up to a relative error of less than \percent{0.4} for both meshes and \gls*{cfl} numbers. 
For $Ra = \numINT{218000}$, a maximum relative error of less than \percent{1} for all extrema is reached for the finer mesh with $48 \times 32$ squares.
We conclude that the computed results agree well with those reported in \cite{Blankenbach:1989:GJI,Vynnytska:2013:C+G}.
The characteristic trends of $\mathbf{u}_\text{rms}$ and $\text{Nu}$ for both scenarios, with mesh size $48 \times 32$ and $\text{CFL} = 1.0$ are plotted in 
\cref{fig:blankenbach-plots}.

\begin{figure}[!ht]
  \footnotesize
  \centering
  \begin{subfigure}[t]{0.48\textwidth}
    \resizebox*{\textwidth}{!}{%
      \input{figures/ra_216000_cfl_1p0_level_3_nx_6.db.pgf} 
    }
    \caption{Ra = \numINT{216000}}
  \end{subfigure}
  \hfill
  \begin{subfigure}[t]{0.48\textwidth}
    \resizebox*{\textwidth}{!}{%
      \input{figures/ra_218000_cfl_1p0_level_3_nx_6.db.pgf}
    }
    \caption{Ra = \numINT{218000}}
  \end{subfigure}
	\caption{$\mathbf{u}_\text{rms}$ and $\text{Nu}$ plotted over a $P2$-cycle for $\text{Ra} = \numINT{216000}$, and a $P4$-cycle for 
	$\text{Ra} = \numINT{218000}$ (mesh: $48 \times 32$, CFL = 1).}
    \label{fig:blankenbach-plots}
\end{figure}

\subsection{Mantle convection on a spherical shell}\label{sec:mc}

As a demonstrator for the applicability to large scale applications, we employ the coupled method to
simulate isoviscous convection with $\text{Ra} = 10^8$ and no internal heating ($q = 0$).
The domain approximates Earth's mantle by the spherical shell
$\Omega = \{ \mathbf{x} \in \mathbb{R}^3 : r_\text{min} \leq \norm{\mathbf{x}}_2 \leq r_\text{max} \}$
with $r_\text{min} = 0.5$ and $r_\text{max} = 1$.
The computational grid is composed of \numINT{19200} tetrahedral macro-cells, which are refined $4$ times
and projected onto the sphere, resulting in more than \num{3.2e8} unknowns for the Stokes equation,
and \num{1.0e8} \glspl*{dof} (and therefore particles) for the advection-diffusion equation, solved for
in every time step. 

The initial and Dirichlet boundary conditions for the temperature are prescribed by
$c_0(r) = \exp\left(-10 \frac{r - r_\text{min}}{r_\text{max} - r_\text{min}}\right)$
where $r$ is the distance to the origin.
For the velocity, we set no-slip boundary conditions at all boundaries.

We apply the predictor-corrector scheme in \cref{alg:pc} with Strang-splitting, and
simulate \numINT{3000} time-steps with a \gls*{cfl}-number of $1$.
The Stokes system is solved with a monolithic geometric multigrid solver that employs an inexact
Uzawa smoother with weighted Jacobi relaxation \cite{Kohl:2020:arXiv}.
Its excellent performance and scalability to linear systems more than a trillion ($10^{12}$) unknowns is 
discussed in \cite{Gmeiner:2015:SISC,Gmeiner:2016:JoCS,Kohl:2020:arXiv}.
For the diffusive term, \ie the solution of the linear system \cref{eq:linear-system},
we employ a standard conjugate gradient iteration, which turns out to be sufficient.

\begin{figure}[!ht]
  \footnotesize
  \centering
  \subcaptionbox{parameter summary\label{fig:mc-details-parameters}}{%
      \centering
      \resizebox*{0.48\textwidth}{!}{
      \begin{tabular}{ll}
        \multicolumn{2}{c}{Mantle convection benchmark parameters} \\
        \midrule
        machine & \supermucng \\
        nodes & \numINT{400} \\
        cores & \numINT{19200} \\
        \addlinespace
        numerical scheme & predictor-corrector (see \cref{alg:pc}) \\
        \addlinespace
        Stokes & \\
        \hspace{6pt} discretization & \Ptwo-\Pone (Taylor-Hood) \\
        \hspace{6pt} solver & \makecell[tl]{monolithic geometric multigrid (GMG)} \\
        \hspace{6pt} \glspl*{dof} & \num{3.2e8} \\
        \addlinespace
        temperature & \\
        \hspace{6pt} discretization & \Ptwo finite elements + \gls*{mmoc} \\
        \hspace{6pt} advection scheme & \gls*{mmoc} \\
        \hspace{6pt} solver diffusion & conjugate gradient (CG) \\
        \hspace{6pt} \glspl*{dof} & \num{1.0e8} \\
        \addlinespace
        Rayleigh number & $10^8$ \\
        \gls*{cfl} & \numINT{1} \\
        avg. run time / ts & $\approx 19$s (incl. pred. + corr., I/O)\\
      \end{tabular}
    }}
  \hfill
  \subcaptionbox{run time of components of \cref{alg:pc}\label{fig:mc-details-stackplot}}{
    \includegraphics[trim=65 1 1 40,clip,width=0.48\textwidth]{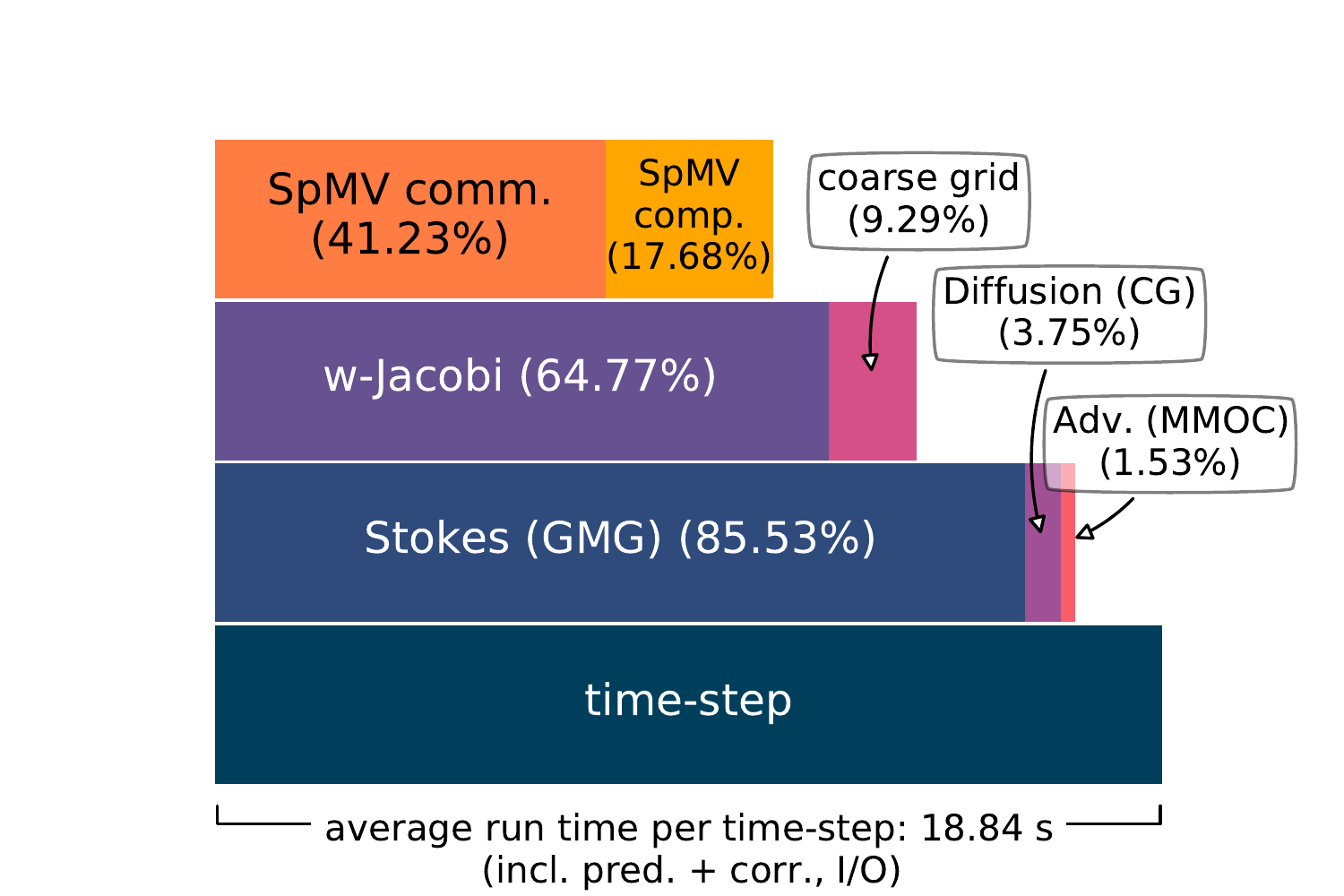} 
  }
  \caption{Mantle convection benchmark: (\protect\subref{fig:mc-details-parameters}) Summary of parameters. 
  (\protect\subref{fig:mc-details-stackplot}) Stacked bar chart of average fractional run time of components of \cref{alg:pc}.
  The percentage in parentheses indicates the fractional run time with respect to the overall run time of a predictor-corrector step.}
  \label{fig:mc-details}
\end{figure}

The simulation is performed on 400 nodes (\numINT{19200} processes) of \supermucng in roughly 16 hours.
In \cref{fig:mc-details-parameters} we list a summary of the benchmark parameters.
\Cref{fig:mc-details-stackplot} shows a stacked bar chart of the fractional run time of the relevant
components of the predictor-corrector scheme.
On average, the computation of a single time-step takes about \numINT{19} seconds. 
About \percent{85} of the total run time is spent for the solution of the Stokes system.
Almost half of that time (roughly \percent{41} of total run time) accounts for communication
during the Jacobi-relaxation. Especially during iterations on the coarser grids, communication time
strongly dominates time spent in the compute kernels. Strategies to further improve the performance of the
coarse grid solver are presented in \cite{Buttari:2020:Block}.

\begin{figure}[!ht]
  \footnotesize
  \centering
  \begin{subfigure}[t]{0.48\textwidth}
    \includegraphics[width=\textwidth]{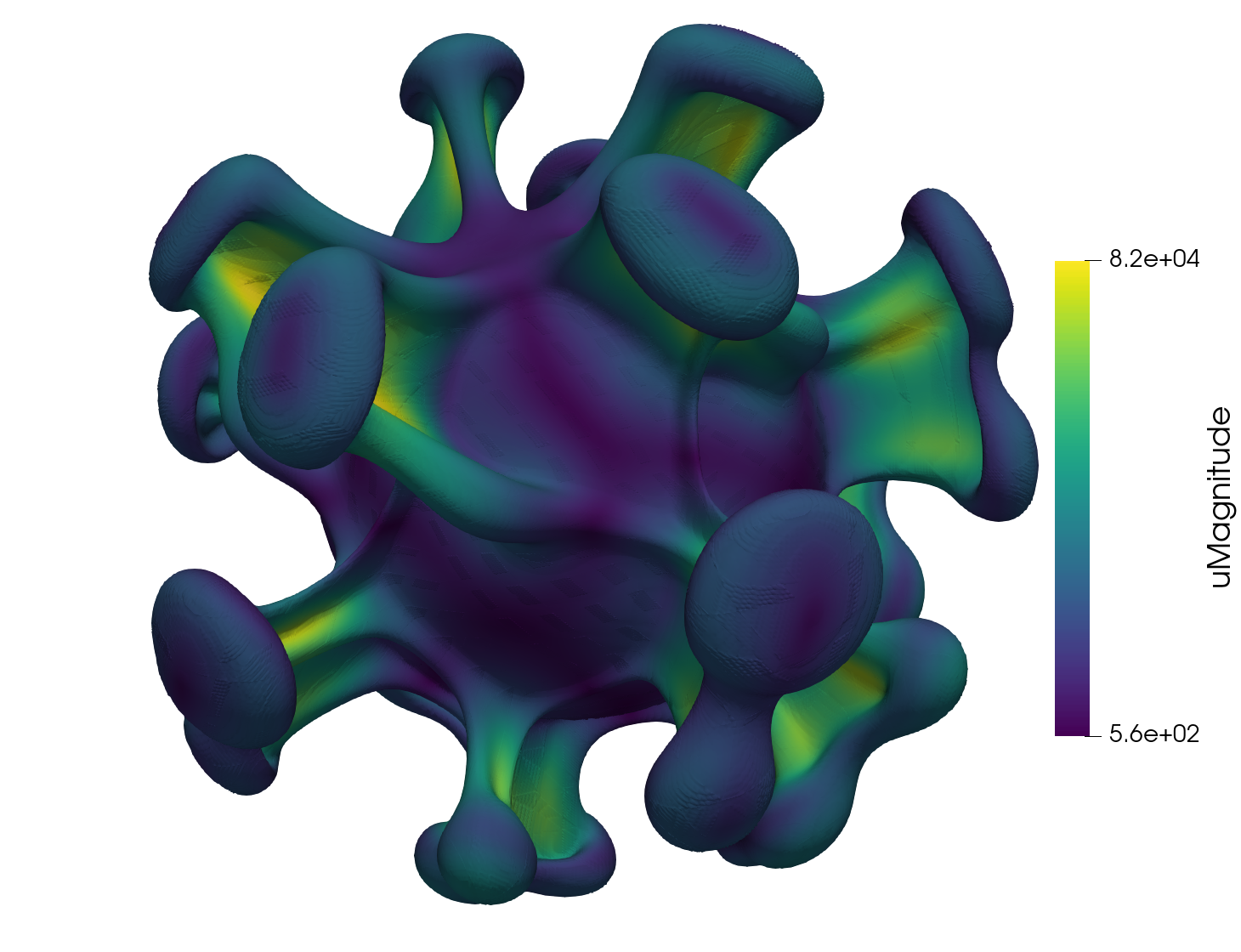}
    \caption{time-step 200}
  \end{subfigure}
  \hfill
  \begin{subfigure}[t]{0.48\textwidth}
    \includegraphics[width=\textwidth]{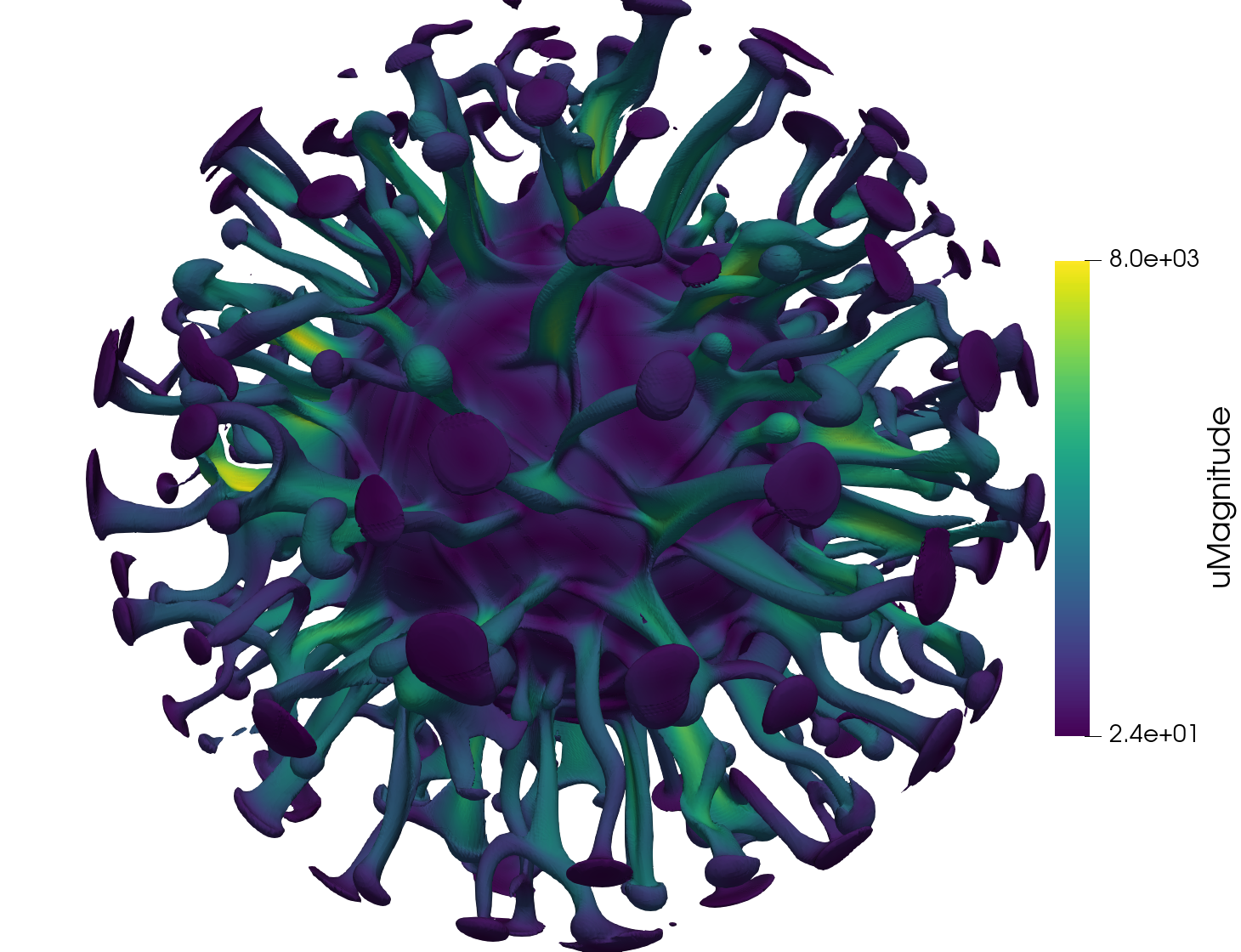}
    \caption{time-step 3000}
  \end{subfigure}
  \caption{Contour plot at $c = 0.15$ of the temperature solution on the spherical shell for
    $\text{Ra} = 10^8$, colored by velocity magnitude.}
  \label{fig:mc}
\end{figure}

In \cref{fig:mc}, the contour surfaces of the temperature at $c_\text{cont} = 0.15$ are shown
at time steps \# \numINT{200}, and \numINT{3000}.
Thin, chaotically rising plumes are observed as expected at such large Rayleigh numbers.

\section*{Conclusion}

In this article, we presented an implementation of an Eulerian-Lagrangian discretization
based on the method of characteristics to treat the advection-diffusion equation in the
advection-dominated regime. Its numerical performance was demonstrated
on multiple two- and three-dimensional benchmarks, including cases with pure advection,
curved geometries, and discontinuous solutions.

Motivated by the demand of extreme spatial resolution in mantle convection simulations, 
the parallel scalability of our implementation was assessed in a weak and strong scaling 
benchmark for the advection-diffusion equation.
We demonstrate a parallel efficiency of more than \percent{92}, solving for more than \num{5.2e10} 
\glspl*{dof} per time-step on \numINT{147456} parallel processes.
Finally, we applied the method to buoyancy-driven Stokes flow, embedding it into a 
non-linear scheme based on a predictor-corrector method. The scheme was verified through
a classical benchmark for time-dependent convection, and its practical applicability to
large scale problems demonstrated in a mantle convection benchmark on the spherical shell,
with combined more than \num{4.0e8} unknowns solved for in each time step for
\numINT{3000} time steps.

\section*{Acknowledgements}

The authors gratefully acknowledge the Gauss Centre for Supercomputing e.V. (\url{www.gauss-centre.eu}) for funding 
this project by providing computing time on the GCS Supercomputer SuperMUC-NG at Leibniz Supercomputing Centre (\url{www.lrz.de}).
The authors also gratefully acknowledge financial support by the Bavarian State Ministry of Science and the Arts through the Competence 
Network for Scientific High Performance Computing in Bavaria (KONWIHR) and by the German Research Foundation through the Priority Programme 
1648 Software for Exascale Computing (SPPEXA), RU 422/16-2.

\bibliographystyle{plain}
\bibliography{bibliography-additional}

\begin{thebibliography}{10}

\bibitem{Allievi:2000:IJNMF}
Alejandro Allievi and Rodolfo Bermejo.
\newblock Finite element modified method of characteristics for the
  {N}avier-{S}tokes equations.
\newblock {\em Int.~J.~Numer.~Meth.~Fluids}, 32(4):439--463, 2000.

\bibitem{Bauer:2020:SPPEXA}
S.~Bauer, H.-P. Bunge, D.~Drzisga, S.~Ghelichkhan, M.~Huber, N.~Kohl, M.~Mohr,
  U.~R{\"u}de, D.~Th{\"o}nnes, and B.~Wohlmuth.
\newblock {TerraNeo} --- {M}antle {C}onvection {B}eyond a {T}rillion {D}egrees
  of {F}reedom.
\newblock In H.-J. Bungartz, S.~Reiz, B.~Uekermann, P.~Neumann, and W.~Nagel,
  editors, {\em Software for Exascale Computing - SPPEXA 2016-2019}, volume 136
  of {\em Lecture Notes in Computational Science and Engineering}, pages
  569--610. Springer, 2020.

\bibitem{Bauer:2018:SISC}
S.~Bauer, D.~Drzisga, M.~Mohr, U.~R{\"u}de, C.~Waluga, and B.~Wohlmuth.
\newblock A stencil scaling approach for accelerating matrix-free finite
  element implementations.
\newblock {\em SIAM J.~Sci.~Comp.}, 40(6):C748--C778, 2018.

\bibitem{Bauer:2019:JoCS}
S.~Bauer, M.~Huber, S.~Ghelichkhan, M.~Mohr, U.~R{\"u}de, and B.~Wohlmuth.
\newblock Large-scale {S}imulation of {M}antle {C}onvection {B}ased on a {N}ew
  {M}atrix-{F}ree {A}pproach.
\newblock {\em J.~Comput.~Sci.}, 31:60--76, 2019.

\bibitem{Bauer:2017:ANM}
S.~Bauer, M.~Mohr, U.~R{\"u}de, J.~Weism{\"u}ller, M.~Wittmann, and
  B.~Wohlmuth.
\newblock A two-scale approach for efficient on-the-fly operator assembly in
  massively parallel high performance multigrid codes.
\newblock {\em Appl.~Numer.~Math.}, 122:14--38, 2017.

\bibitem{Bergen:2004:NLAA}
B.~Bergen and F.~H{\"u}lsemann.
\newblock Hierarchical hybrid grids: data structures and core algorithms for
  multigrid.
\newblock {\em Numer. Linear Algebra Appl.}, 11:279--291, 2004.

\bibitem{Bermudez:2006pt1:SINUM}
Alfredo Berm{\'u}dez, Maria~R Nogueiras, and Carlos V{\'a}zquez.
\newblock Numerical analysis of convection-diffusion-reaction problems with
  higher order characteristics/finite elements. part i: time discretization.
\newblock {\em SIAM J.~Numer.~Anal.}, 44(5):1829--1853, 2006.

\bibitem{Bey:1995:Tetrahedral}
J{\"u}rgen Bey.
\newblock Tetrahedral grid refinement.
\newblock {\em Computing}, 55(4):355--378, 1995.

\bibitem{Blankenbach:1989:GJI}
B.~Blankenbach, F.~Busse, U.~Christensen, L.~Cserepes, D.~Gunkel, U.~Hansen,
  H.~Harder, G.~Jarvis, M.~Koch, G.~Marquart, D.~Moore, P.~Olson, H.~Schmeling,
  and T.~Schnaubelt.
\newblock A benchmark comparison for mantle convection codes.
\newblock {\em Geophys.~J.~Int.}, 98(1):23--38, 1989.

\bibitem{Brooks:1982:CMAME}
Alexander~N. Brooks and Thomas J.~R. Hughes.
\newblock Streamline upwind/{P}etrov-{G}alerkin formulations for convection
  dominated flows with particular emphasis on the incompressible
  {N}avier-{S}tokes equations.
\newblock {\em Comp.~Meth.~Appl.~Mech.~Engrg.}, 32(1-3):199--259, 1982.

\bibitem{Burstedde:2013:GJI}
Carsten Burstedde, Georg Stadler, Laura Alisic, Lucas~C. Wilcox, Eh~Tan,
  Michael Gurnis, and Omar Ghattas.
\newblock Large-scale adaptive mantle convection simulation.
\newblock {\em Geophys.~J.~Int.}, 192(3):889--906, 2013.

\bibitem{Busto:2020:CAF}
Saray Busto, Maurizio Tavelli, Walter Boscheri, and Michael Dumbser.
\newblock Efficient high order accurate staggered semi-implicit discontinuous
  {G}alerkin methods for natural convection problems.
\newblock {\em Computers \& Fluids}, 198, 2020.

\bibitem{Buttari:2020:Block}
Alfredo Buttari, Markus Huber, Philippe Leleux, Th{\'e}o Mary, Ulrich Ruede,
  and Barbara Wohlmuth.
\newblock {Block Low Rank Single Precision Coarse Grid Solvers for Extreme
  Scale Multigrid Methods}.
\newblock Apr 2020.
\newblock Submitted.

\bibitem{Celia:1990:AWR}
Michael~A. Celia, Thomas~F. Russell, Ismael Herrera, and Richard~E. Ewing.
\newblock An {E}ulerian-{L}agrangian localized adjoint method for the
  advection-diffusion equation.
\newblock {\em Adv.~Water Resources}, 13(4):187--206, 1990.

\bibitem{Chen:2006:SIAM}
Zhangxin Chen, Guanren Huan, and Yuanle Ma.
\newblock {\em Computational {M}ethods for {M}ultiphase {F}lows in {P}orous
  {M}edia}.
\newblock SIAM, 2006.

\bibitem{Cockburn:1998:SINUM}
Bernardo Cockburn and Chi-Wang Shu.
\newblock The {L}ocal {D}iscontinuous {G}alerkin {M}ethod for
  {T}ime-{D}ependent {C}onvection-{D}iffusion {S}ystems.
\newblock {\em SIAM J.~Numer.~Anal.}, 35(6):2440--2463, 1998.

\bibitem{Dawson:1989:SINUM}
CN~Dawson, TF~Russell, and MF~Wheeler.
\newblock Some improved error estimates for the modified method of
  characteristics.
\newblock {\em SIAM J.~Numer.~Anal.}, 26(6):1487--1512, 1989.

\bibitem{Douglas:1999:NUMA}
Jim Douglas, Jr., Chieh-Sen Huang, and Felipe Pereira.
\newblock The modified method of characteristics with adjusted advection.
\newblock {\em Numer.~Math.}, 83(3):353--369, 1999.

\bibitem{Douglas:1982:SINUM}
Jim Douglas, Jr. and Thomas~F. Russell.
\newblock Numerical {M}ethods for {C}onvection-{D}ominated {D}iffusion
  {P}roblems {B}ased on {C}ombining the {M}ethod of {C}haracteristics with
  {F}inite {E}lement or {F}inite {D}ifference {P}rocedures.
\newblock {\em SIAM J.~Numer.~Anal.}, 19(5):871--885, 1982.

\bibitem{Eibl:2019:arXiv}
Sebastian Eibl and Ulrich R{\"u}de.
\newblock A {M}odular and {E}xtensible {S}oftware {A}rchitecture for {P}article
  {D}ynamics.
\newblock Submitted.

\bibitem{Eibl:2018:PARCO}
Sebastian Eibl and Ulrich R{\"u}de.
\newblock A local parallel communication algorithm for polydisperse rigid body
  dynamics.
\newblock {\em Parallel Comput.}, 80:36--48, 2018.

\bibitem{ElAmrani:2008:IJCM}
Mofdi El-Amrani and Mohammed Sea{\"i}d.
\newblock Eulerian-{L}agrangian time-stepping methods for convection-dominated
  problems.
\newblock {\em Int.~J.~Comput.~Math.}, 85(3-4):421--439, 2008.

\bibitem{Elman:2014:OUP}
Howard~C Elman, David~J Silvester, and Andrew~J Wathen.
\newblock {\em Finite {E}lements and {F}ast {I}terative {S}olvers with
  {A}pplications in {I}ncompressible {F}luid {D}ynamics}.
\newblock Oxford University Press, 2nd edition, 2014.

\bibitem{Gassmoeller:2019:GJI}
R.~Gassm\"{o}ller, H.~Lokavarapu, W.~Bangerth, and E.~G. Puckett.
\newblock Evaluating the accuracy of hybrid finite element/particle-in-cell
  methods for modelling incompressible {S}tokes flow.
\newblock {\em Geophys.~J.~Int.}, 219(3):1915--1938, 2019.

\bibitem{Gmeiner:2015:SISC}
B.~Gmeiner, U.~R{\"u}de, H.~Stengel, C.~Waluga, and B.~Wohlmuth.
\newblock Performance and scalability of hierarchical hybrid multigrid solvers
  for stokes systems.
\newblock {\em SIAM J.~Sci.~Comput.}, 37(2):C143--C168, 2015.

\bibitem{Gmeiner:2016:JoCS}
Bj{\"o}rn Gmeiner, Markus Huber, Lorenz John, Ulrich R{\"u}de, and Barbara
  Wohlmuth.
\newblock A quantitative performance study for {S}tokes solvers at the extreme
  scale.
\newblock {\em J.~Comput.~Sci.}, 17(3):509--521, 2016.

\bibitem{Gordon:1973:NUMA}
William~J Gordon and Charles~A Hall.
\newblock Transfinite {E}lement {M}ethods: {B}lending-{F}unction
  {I}nterpolation over {A}rbitary {C}urved {E}lement {D}omains.
\newblock {\em Numer.~Math.}, 21(2):109--129, 1973.

\bibitem{John:2008:CMAME}
Volker John and Ellen Schmeyer.
\newblock Finite element methods for time-dependent
  convection-diffusion-reaction equations with small diffusion.
\newblock {\em Comp.~Meth.~Appl.~Mech.~Engrg.}, 198(3-4):475--494, 2008.

\bibitem{Kohl:2020:arXiv}
Nils Kohl and Ulrich R{\"u}de.
\newblock Textbook efficiency: massively parallel matrix-free multigrid for the
  {S}tokes system, 2020.
\newblock Submitted.

\bibitem{Kohl:2019:IJPEDS}
Nils Kohl, Dominik Th\"onnes, Daniel Drzisga, Dominik Bartuschat, and Ulrich
  R\"ude.
\newblock The {HyTeG} finite-element software framework for scalable multigrid
  solvers.
\newblock {\em Int.~J.~Par., Emerg.~Distrib.~Sys.}, 34(5):477--496, 2019.

\bibitem{Kronbichler:2012:GJI}
M.~Kronbichler, T.~Heister, and W.~Bangerth.
\newblock High {A}ccuracy {M}antle {C}onvection {S}imulation through {M}odern
  {N}umerical {M}ethods.
\newblock {\em Geophys.~J.~Int.}, 191(1):12--29, 2012.

\bibitem{Kuzmin:2004:CMAME}
Dimitri Kuzmin, Matthias M{\"o}ller, and Stefan Turek.
\newblock High-resolution {FEM}--{FCT} schemes for multidimensional
  conservation laws.
\newblock {\em Comput.~Methods Appl.~Mech.~Engrg.}, 193(45-47):4915--4946,
  2004.

\bibitem{Kuzmin:2012:Springer}
Dmitri Kuzmin.
\newblock Algebraic {F}lux {C}orrection {I} - {C}onservation {L}aws.
\newblock In {\em Flux-{C}orrected Transport - Principles, Algorithms, and
  Applications}. Springer, 2012.

\bibitem{LeVeque:1996:SINUM}
Randall~J. LeVeque.
\newblock High-{R}esolution {C}onservative {A}lgorithms for {A}dvection in
  {I}ncompressible {F}low.
\newblock {\em SIAM J.~Numer.~Anal.}, 33(2):627–665, 1996.

\bibitem{Malevsky:1991:PFA}
A.~V. Malevsky and D.~A. Yuen.
\newblock Characteristics-based methods applied to infinite {P}randtl number
  thermal convection in the hard turbulent regime.
\newblock {\em Phys.~Fluids A}, 3(9):2105--2115, 1991.

\bibitem{Morton:2019:CRC}
Keith~W. Morton.
\newblock {\em Numerical {S}olution of {C}onvection-{D}iffusion {P}roblems}.
\newblock CRC Press, 2019.

\bibitem{Ouro:2019:CAF}
Pablo Ouro, Bru\ {n}o Fraga, and Unai Lopez-Novoac amd Thorsten~Stoesser.
\newblock Scalability of an {E}ulerian-{L}agrangian large-eddy simulation
  solver with hybrid {MPI}/{OpenMP} parallelisation.
\newblock {\em Computers \& Fluids}, 179:123--136, 2019.

\bibitem{Quarteroni:2008:Springer}
Alfioc Quarteroni and Alberto Valli.
\newblock {\em Numerical {A}pproximation of {P}artial {D}ifferential
  {E}quations}.
\newblock Springer, 2008.

\bibitem{Reinarz:2020:CPC}
Anne Reinarz, Dominic~E Charrier, Michael Bader, Luke Bovard, Michael Dumbser,
  Kenneth Duru, Francesco Fambri, Alice-Agnes Gabriel, Jean-Matthieu Gallard,
  Sven K{\"o}ppel, et~al.
\newblock {ExaHyPE: an engine for parallel dynamically adaptive simulations of
  wave problems}.
\newblock {\em Computer Physics Communications}, 254:107251, 2020.

\bibitem{Ricard:2007:Treatise}
Y.~Ricard.
\newblock Physics of {M}antle {C}onvection.
\newblock In David Bercovici, editor, {\em Mantle Dynamics}, volume~7 of {\em
  Treatise on Geophysics}, pages 31--89. Elsevier, 2007.

\bibitem{Rudi:2015:SC}
Johann Rudi, A.~Cristiano~I. Malossi, Tobin Isaac, Georg Stadler, Michael
  Gurnis, Peter W.~J. Staar, Yves Ineichen, Costas Bekas, Alessandro Curioni,
  and Omar Ghattas.
\newblock {An Extreme-Scale Implicit Solver for Complex PDEs: Highly
  Heterogeneous Flow in Earth’s Mantle}.
\newblock In {\em Proceedings of the International Conference for High
  Performance Computing, Networking, Storage and Analysis}, SC '15, pages
  5:1--5:12. ACM, 2015.

\bibitem{Russell:2002:AWR}
Thomas~F. Russell and Michael~A. Celia.
\newblock An overview of research on {E}ulerian--{L}agrangian localized adjoint
  methods {(ELLAM)}.
\newblock {\em Adv.~Water Resources}, 25(8-12):1215--1231, 2002.

\bibitem{Samake:2017:JCP}
Abdoulaye Samak\'{e}, Pierre Rampal, Sylvain Bouillon, and Einar \'{O}lason.
\newblock Parallel implementation of a {L}agrangian-based model on an adaptive
  mesh in {C++}: {A}pplication to sea-ice.
\newblock {\em J,~Comput.~Phys.}, 350:84--96, 2017.

\bibitem{Strang:1968:SINUM}
Gilbert Strang.
\newblock On the construction and comparison of difference schemes.
\newblock {\em SIAM J.~Numer.~Anal.}, 5(3):506--517, 1968.

\bibitem{Tavelli:2019:IJNMF}
Maurizio Tavelli and Walter Boscheri.
\newblock A high‐order parallel {E}ulerian-{L}agrangian algorithm for
  advection-diffusion problems on unstructured meshes.
\newblock {\em Int.~J.~Numer.~Meth.~Fluids}, 91:332--347, 2019.

\bibitem{HyTeG:2021:SW}
{TerraNeo Team}.
\newblock \gls*{hyteg}, 2021.
\newblock SHA: 3e5a93130cc9b86143cfa5249351914ee99cd6e3.

\bibitem{Trottenberg:2001:GreyBook}
Ulrich Trottenberg, Cornelius Oosterlee, and Anton Sch{\"u}ller.
\newblock {\em Multigrid}.
\newblock Academic Press, 2001.

\bibitem{vandenBerg:1993:GJI}
Arie~P. van~den Berg, Peter~E. van Keken, and David~A. Yuen.
\newblock The effects of a composite non-{N}ewtonian and {N}ewtonian rheology
  on mantle convection.
\newblock {\em Geophys.~J.~Int.}, 115(1):62--78, 1993.

\bibitem{Vynnytska:2013:C+G}
L.~Vynnytska, M.~E. Rognes, and S.~R. Clark.
\newblock Benchmarking {FEniCS} for mantle convection simulations.
\newblock {\em Comp.~\& Geosci.}, 50:95--105, 2013.

\bibitem{Waluga:2016:JCP}
Christian Waluga, Barbara Wohlmuth, and Ulrich R{\"u}de.
\newblock Mass-corrections for the conservative coupling of flow and transport
  on collocated meshes.
\newblock {\em J.~Comput.~Phys.}, 305:319--332, 2016.

\bibitem{Zalesak:1979:JCP}
Steven~T. Zalesak.
\newblock Fully multidimensional flux-corrected transport algorithms for
  fluids.
\newblock {\em J.~Comput.~Phys.}, 31(3):335--362, 1979.

\end{thebibliography}

\end{document}